\documentclass[preprint2]{aastex}
\usepackage{graphicx,float}
\usepackage{amsmath}
\usepackage{natbib}
\usepackage{aas_macros}
\usepackage{enumerate}
\usepackage{amssymb}
\usepackage{url}
\usepackage[colorlinks=true, pdfstartview=FitV, linkcolor=blue, citecolor=blue, urlcolor=blue]{hyperref} 

\begin{document}

\title{An Interacting Galaxy System Along a Filament in a Void}

\author{B. Beygu\altaffilmark{1}, K. Kreckel\altaffilmark{2}, R. van de Weygaert\altaffilmark{1}, J. M. van der Hulst\altaffilmark{1}, J. H. van Gorkom\altaffilmark{3} }

\altaffiltext{1}{Kapteyn Astronomical Institute, University of Groningen, PO Box 800, 9700 AV Groningen, the Netherlands}
\altaffiltext{2}{Max Planck Institute for Astronomy, K\"{o}nigstuhl 17, 69117 Heidelberg, Germany}
\altaffiltext{3}{Department of Astronomy, Columbia University, Mail Code 5246, 550 West 120th Street, New York, NY 10027, USA}

\email{beygu@astro.rug.nl}

\begin{abstract}
Cosmological voids provide a unique environment for the study of galaxy formation and evolution. The galaxy population in their interior have 
significantly different properties than average field galaxies. As part of our  Void Galaxy Survey (VGS), we have found 
a system of three interacting galaxies (VGS\_31) inside a large void. VGS\_31 is a small elongated group whose members are embedded in a 
common HI envelope. The HI picture suggests a filamentary structure with accretion of intergalactic cold gas from the filament onto the galaxies.  
We present deep optical and narrow band $\rm{H_{\alpha}}$ data, optical spectroscopy, near-UV and far-UV GALEX and CO(1-0) data. We find that one of 
the galaxies, a Markarian object, has a ring-like structure and a tail evident both in optical and HI. While all three galaxies form stars in 
their central parts, the tail and the ring of the Markarian object are devoid of star formation. We discuss these findings in terms of a 
gravitational interaction and ongoing growth of galaxies out of a filament. VGS\_31 is one of the first observed examples of a filamentary structure
in a void. It is an important prototype for understanding the formation of substructure in a void. This system also shows that the galaxy evolution in 
voids can be as dynamic as in high density environments.
\end{abstract}


\section{Introduction}

Voids are vast regions occupying most of the volume in the universe with sizes in 
the range of 20 - 50$h^{-1}$Mpc, usually roundish in shape and largely devoid of galaxies \citep[see][for a recent review]{weyplaten2011}.  
In the large scale structure of the universe we observe today, the most striking features along with the voids are clusters and filaments. 
In this picture, galaxies are distributed in a filament-dominated web-like structure. Filaments connect clusters to each other and, while tenuous, 
act like bridges \citep{zeldovich1970,shandarin1989,bond1996,colberg2005a,aragon2010b}. From recent redshift surveys like the second Center for Astrophysics Redshift Survey \citep{huchra1983}, 
the 2dF Galaxy Redshift Survey \citep{colless2003}, the Sloan Digital Sky Survey (SDSS) \citep{york2000} and 2MASS redshift survey \citep{huchra2012} we see how  filaments, bridges and sheet-like
structures form substructures and surround the underdense regions.

Notwithstanding the very low density of the void regions, we do find a dilute population of 
galaxies in their interior. These {\it void galaxies}  appear to have significantly different properties than
average field galaxies. Previous studies based on redshift surveys, have shown that the void galaxies are in general small, star forming blue galaxies. They have a later
morphological type and have higher specific star formation rates than the galaxies in average density environment. Largely unaffected by the complexities and processes modifying
galaxies in high-density environments, the galaxies living in the
isolated void regions are expected to have retained vital clues to
their formation and evolution. It has made the study of the relation
between void galaxies and their surroundings an important aspect of
the recent interest in environmental influences on galaxy formation
\citep{szomoru1996,kuhn1997,popescu1997,karachentseva1999,grogin1999,grogin2000,peebles2001,hoyle2002a,hoyle2002b,
rojas2004,rojas2005,tikhonov2006,patiri2006a,patiri2006b,ceccar2006,wegner2008,stanonik2009,kreckel2011,
kreckel2012}.

Void galaxies may be the rare probes of the faint and tenuous substructure that hierarchical structure formation 
theories predict to exist in voids \citep{dubinski1993,weykamp1993,sahni1994,sheth2004,furlanetto2006,einasto2011,aragon2013}. 
Cosmological simulations show how voids are filled by low-density dark matter filaments, creating a network of tenuous substructures 
within their interior  \citep{weykamp1993,gottloeb2003,colberg2005b,springel2006}. This may indicate that the galaxies residing 
in voids are formed along 
these dark matter filaments, given that the simulations reveal that dark matter haloes are forming along them. 
In fact, some earlier observational studies have found indications for such filamentary substructure in voids \citep{szomoru1996,
zitrin2008}. For example, the latter argue that the dwarf galaxies in their local galaxy sample are located on a dark matter filament 
that itself is located in a low galaxy density region and is accreting intergalactic cold gas onto the filament.

In this study, we present the most outstanding example of such a filamentary void galaxy configuration,VGS\_31, which was 
found within the context of the ``Void Galaxy Survey'' (VGS) \citep{kreckel2011,kreckel2012}. 

We are conducting a multiwavelength survey of 60 void galaxies, called ``The Void Galaxy Survey'' (VGS) \citep{kreckel2011}. 
Galaxies in the VGS have been selected from the Sloan Digital Sky Survey Data Release 7 (SDSS DR7) using purely geometric and topological  
techniques. The sample was selected on the basis of galaxy density maps produced by the Delaunay Tessellation Field Estimator 
(DTFE,\cite{schaap2000,weyschaap2009}) and the subsequent application of the Watershed Void Finder (WVF,\cite{platen2007}; for a 
more general application of the watershed transform to the structural analysis of the cosmic web see \cite{aragon2010a}. The 
combination of DTFE maps 
with WVF detected voids allow us to identify the void galaxies from the deepest interior regions of identified voids in the SDSS 
redshift survey. The goal of this survey is to study the galaxy properties in the most underdense and most pristine environments in the universe where the evolution of galaxies 
is expected to progress more slowly and relatively undisturbed. 

Our geometrically selected sample consist of small galaxies, with stellar mass less than $3 \times 10^{10}$ $\rm{ M_{\odot}}$.
Most of these are small, blue star forming disk galaxies and many of them have companions and extended HI disks, which are often
morphologically and kinematically disturbed \citep{kreckel2011,kreckel2012}. 

\begin{figure*}
\begin{minipage}{180mm}
 \begin{center}
  \includegraphics[scale = 0.25]{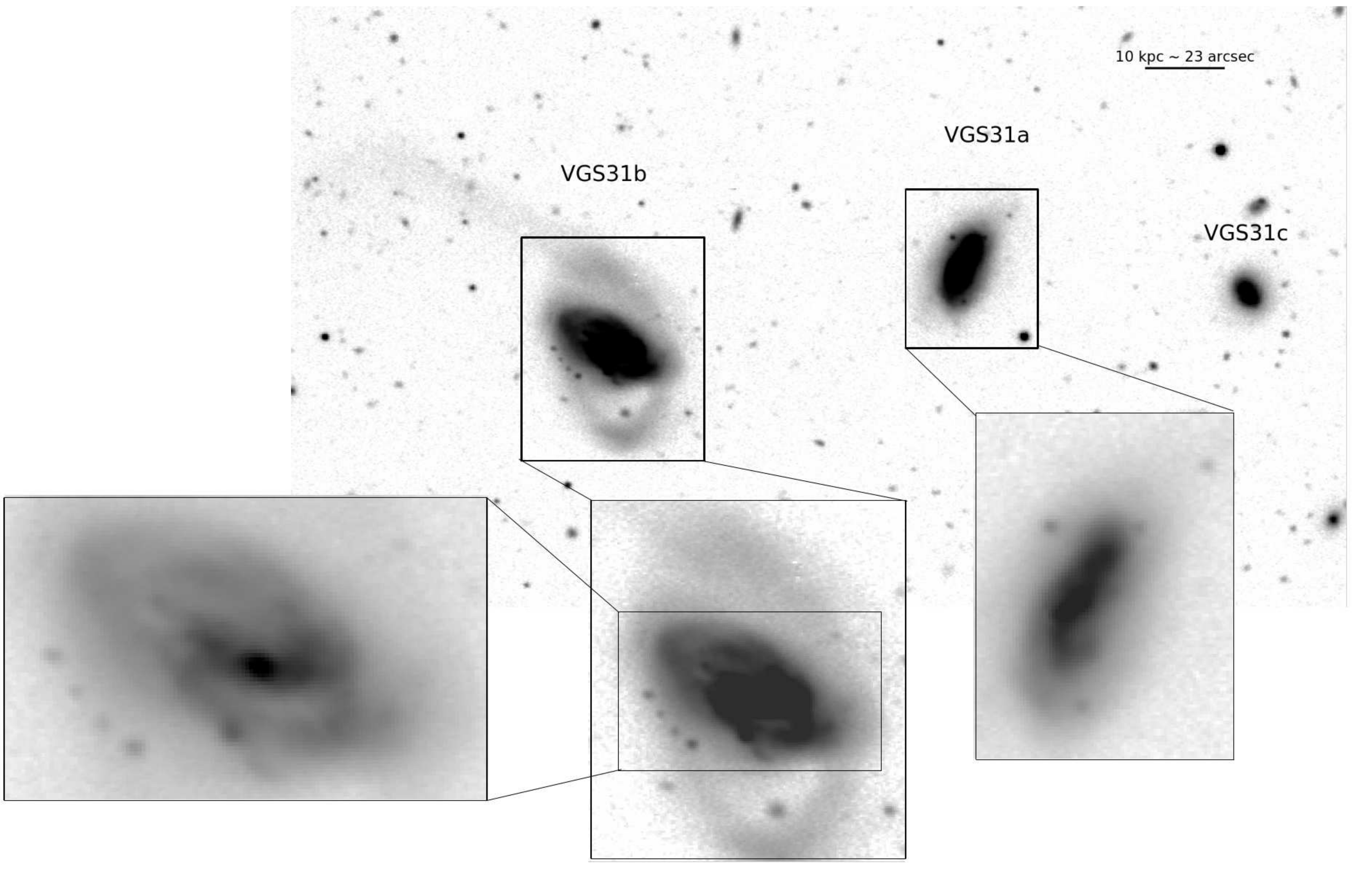} 
  \end{center} 
\end{minipage}
\caption{R band negative image of VGS\_31. From left to right: \textit{VGS\_31b}: The most remarkable member of the system, a \textit{Markarian} galaxy, 
has a tail and a ring. Close up images show the inner structures such as the bar.
\textit{VGS\_31a}: A disk galaxy with a bar structure. \textit{VGS\_31c}:
Smallest member of the system is optically undisturbed. The black bar on the top-right corner represents 10 kpc ($\sim$23$''$).}
 \label{figure:1}
\end{figure*}

\begin{figure*}
 \centering\includegraphics[scale = 0.7]{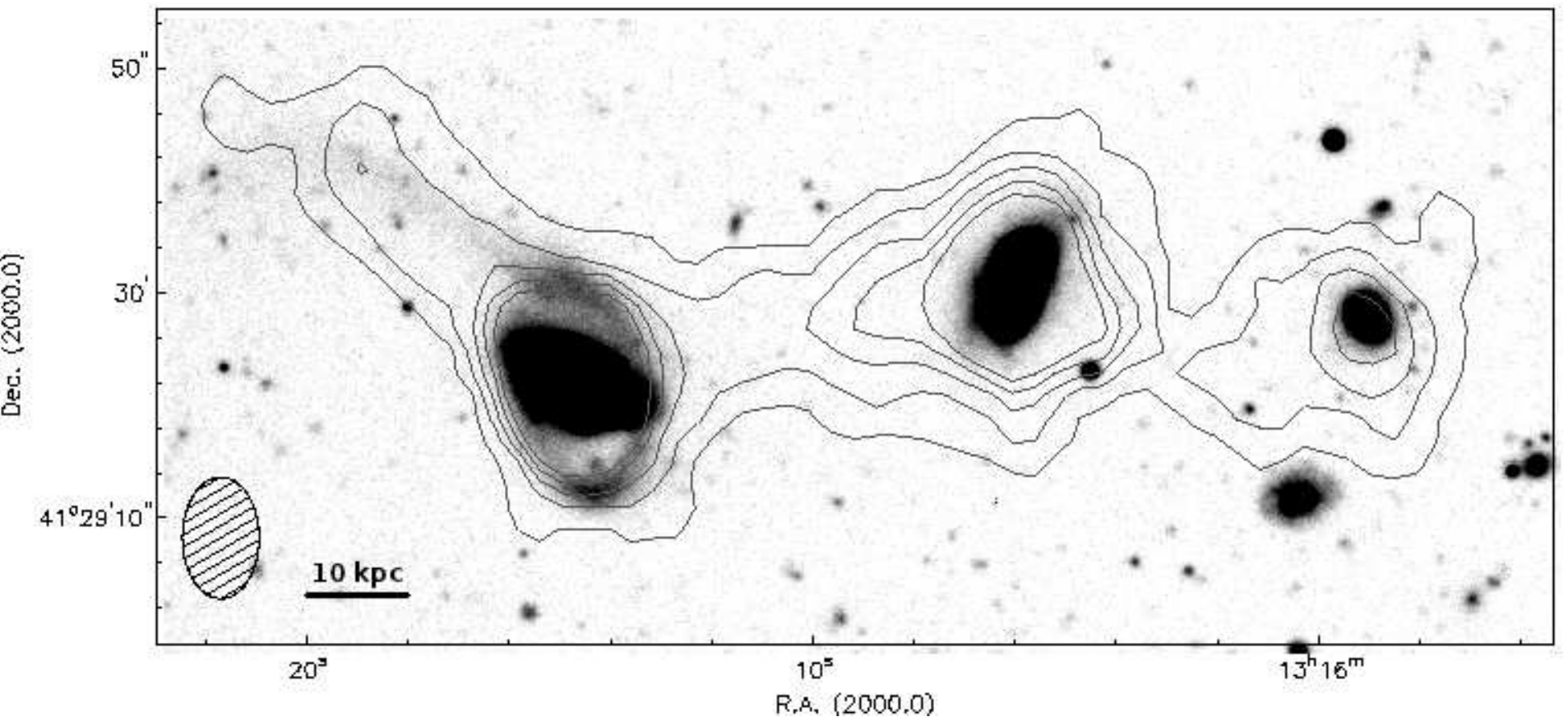} 
  \caption{The HI intensity map of VGS31. HI column density intensity contours start at 1.6 $\times 10^{19} cm^{-2} $ and increase with 4 $\times 10^{19} cm^{-2} $ increments.
Note that members of VGS\_31 are aligned along a HI filament and appear to be embedded in a common HI envelope.}
  \label{figure:3}
\end{figure*}

\begin{figure*}
  \includegraphics[scale = 0.55]{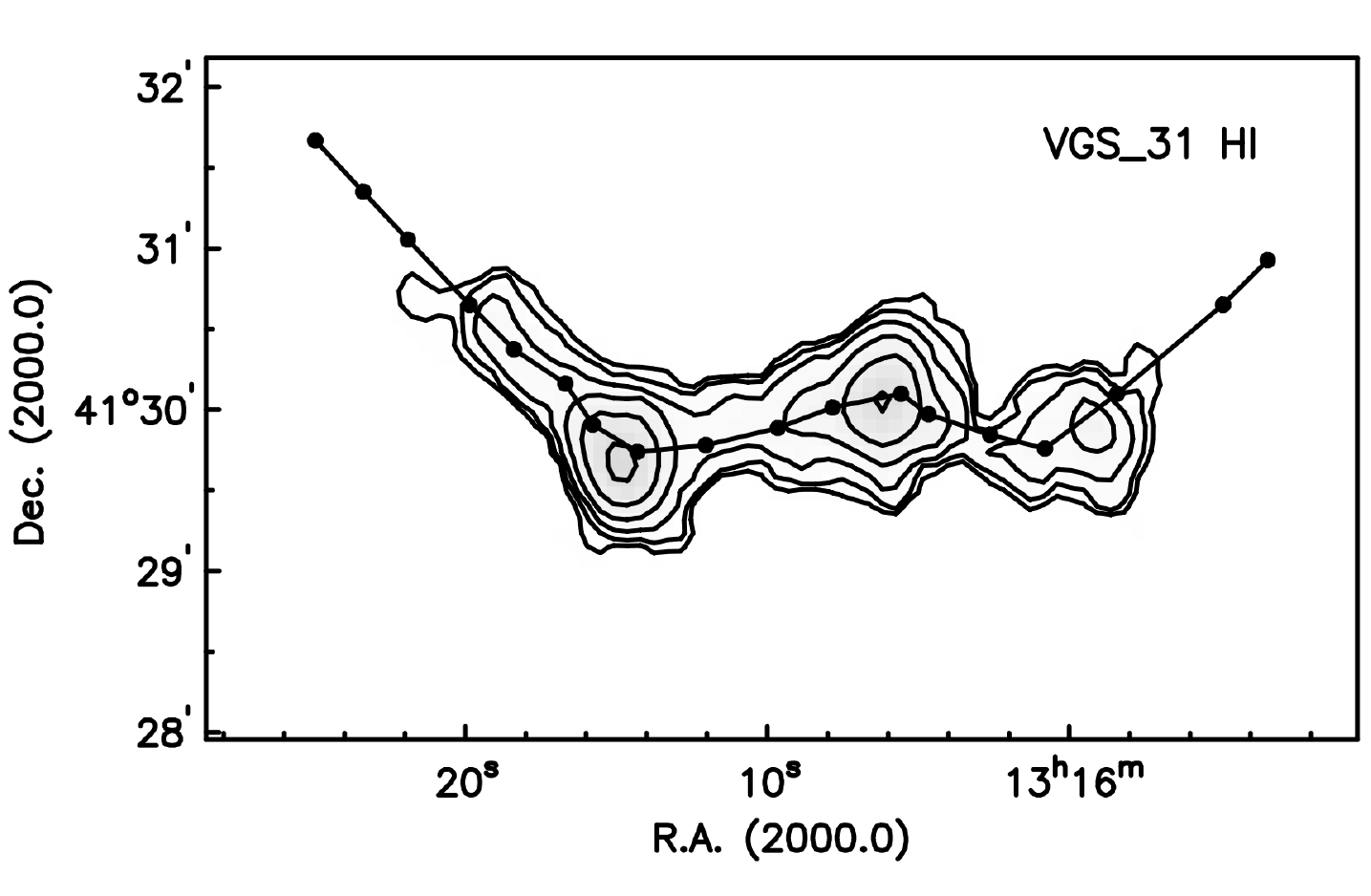}
 \includegraphics[scale = 0.55]{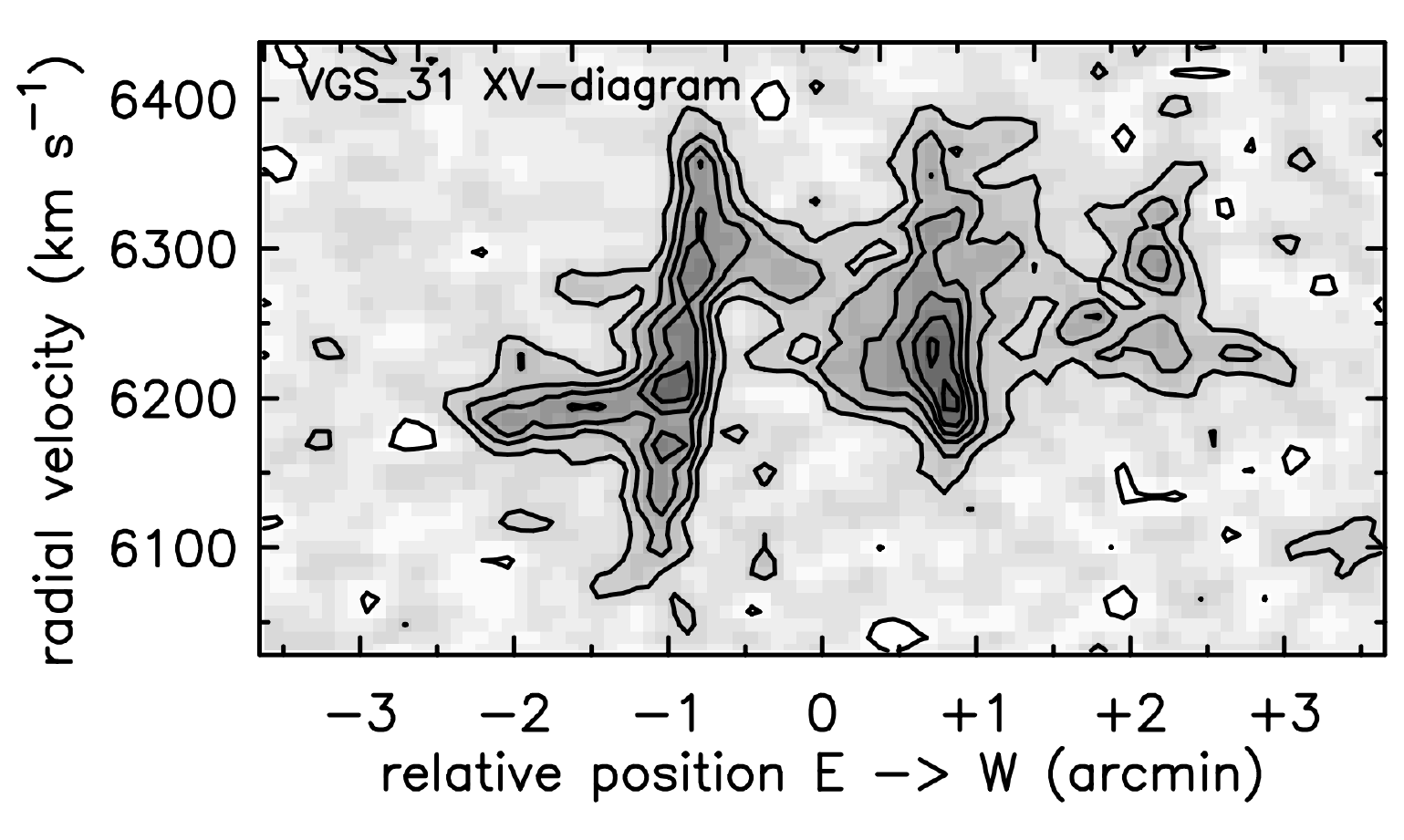}
  \caption{Position velocity (PV) diagram of VGS\_31. Left: Slice position. The black line on the total HI map indicates the position of the slice used to create the position 
velocity (PV) diagram. Black cross is the the zero point (RA: 13:16:10). Right: PV diagram. The PV diagram is created by taking a slice along the whole HI structure, from the beginning of the
tail to the end of VGS\_31c, connecting each points. Zero corresponds to the red dot in the HI image on the left.  This plot shows the velocity structure of the tail and its connection to 
gas around VGS\_31b. Also, it shows the gas between VGS\_31b and VGS\_31a and VGS\_31a and VGS\_31c, respectively. Notice the change in velocity width between VGS\_31b and VGS\_31a.  }
  \label{figure:4}
\end{figure*}

We have found a system of three linearly aligned galaxies, VGS\_31, in a void as part of the
VGS (Figure~\ref{figure:1} and 2). A remarkable feature of VGS\_31 is that the whole system is embedded in a common HI
envelope (Figure~\ref{figure:3}) and the three galaxies are at almost the same velocity (Figure~\ref{figure:4}). The fact that there is a small velocity gradient throughout whole HI
envelope, from the far east of VGS\_31b to far west to VGS\_31c, suggests that this is a filament in which the three galaxies are embedded.

\begin{figure*}
\begin{minipage}{180mm}
 \begin{center}
  \includegraphics[scale = 0.5]{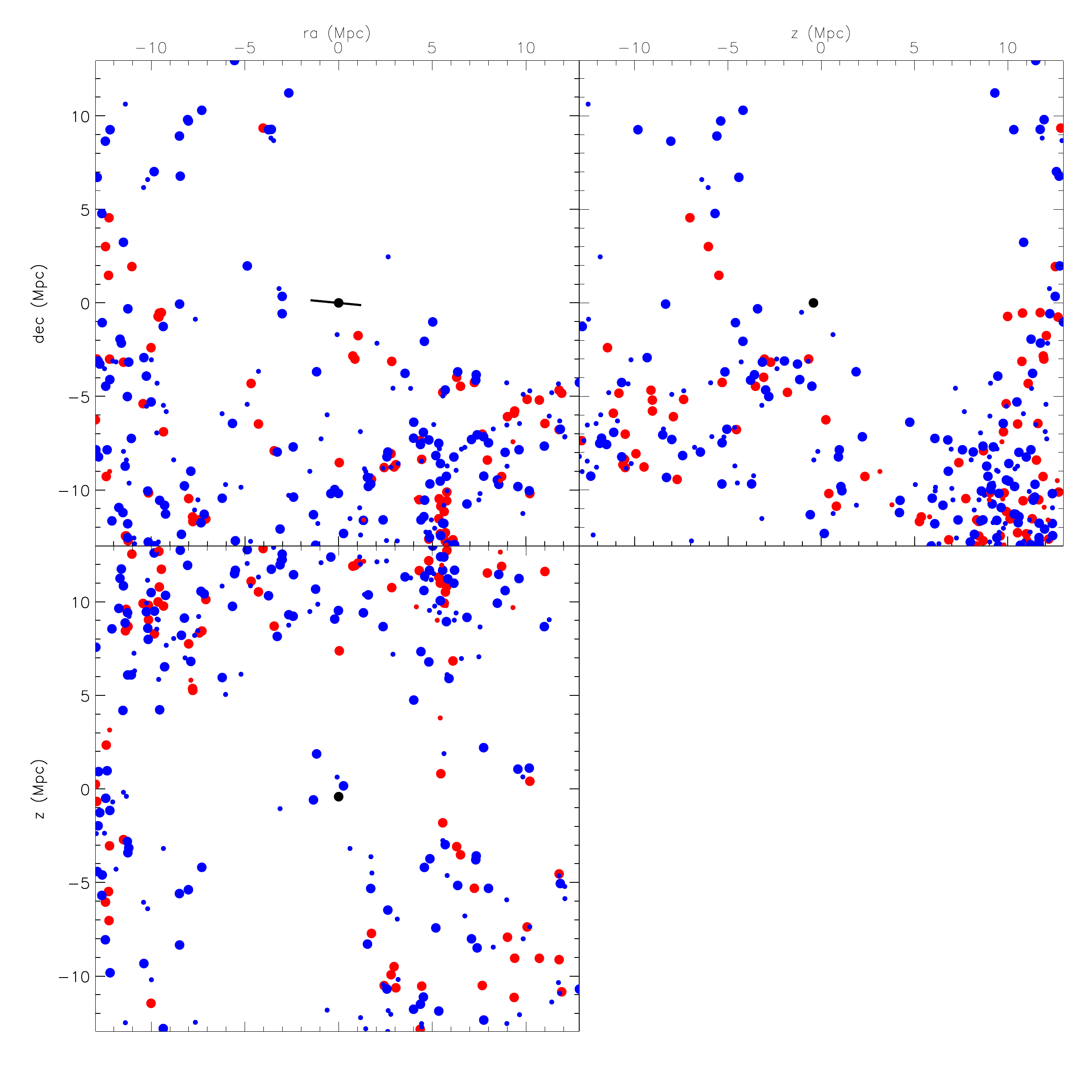} 
  \end{center} 
\end{minipage}
\caption{The large scale structure distribution of galaxies within 13 Mpc around the VGS\_31 system (in black) for different projections in right 
ascension, declination, and redshift space. The line indicates the orientation of the system on the plane of the sky. Surrounding galaxies are color 
coded by g-r color, to be red if g-r$ >$ 0.6 and blue if g-r $\leq$ 0.6. The symbol size indicates luminosity, with larger symbols if $M_{r}$ $<$ -18 and smaller 
symbols if $M_{r}$ $\geq$-18.}
 \label{figure:2}
\end{figure*}

The system exhibits strong signs of interactions (Figure~\ref{figure:1}) and star forming activity with signs of starbursts (Figure~\ref{figure:5} 
and~\ref{figure:6}). VGS\_31 consists of a 
central galaxy VGS\_31a and two companions; VGS\_31b and VGS\_31c. VGS\_31a is optically slightly disturbed. It has a bar like structure and all the star formation activity is concentrated 
there. VGS\_31b is a Markarian galaxy (Mrk 1477). Markarian galaxies are known to have UV continuum excess in their spectra. They have relatively high 
star formation rates and many of them contain
AGNs and starburst nuclei. They span a wide luminosity range between -23 and -13 mag and have a broad range of morphologies. One interesting statistic is 
that an unusual number of Markarian
galaxies occur in tight pairs or interacting systems. In this context, VGS\_31b is a typical example of Markarian galaxy. VGS\_31b
is a starburst galaxy and has enhanced star formation at center of the disk mostly concentrated in the bar.It has a tail, visible both in optical and in 
HI and a ring like structure
around the disk. There is no sign for ongoing star formation activity neither in the tail nor in the ring (Figure~\ref{figure:5}). VGS\_31c is 
significantly smaller than the other two
galaxies and forms stars in its central part as well.

At first glance VGS\_31 looks like a normal interacting system. However the observed properties described above indicate a more complex picture as we may 
be witnessing the growth of structure along a filament within a large cosmological  void. In fact, in an accompanying paper by Rieder et al. 2013 (subm.) 
we have explored the dynamical evolution of the growth of systems resembling VGS\_31 in voids, within the context of the $\Lambda$CDM cosmology. In this 
study we analysed the high-resolution  $\Lambda$CDM simulation Cosmo-Grid \citep{portegies2010} to see how dark matter halo systems similar in mass, size and 
environment to VGS\_31 came to be.  We found eight systems as suitable candidates for harbouring a VGS\_31 like system and then investigated their
merger histories. We found that while VGS31-like systems have a large variation in formation time, the environment in which they are embedded evolved 
very similarly. It seems 
to suggest that we may be witnessing the assembly of a filament in a void by bringing together several smaller filamentary structures, each populated 
by individual haloes. It conjures up the interesting question whether the galaxies in the VGS\_31 configuration were formed at the same location or whether 
they each originate from a different location. 

In order to study the current properties and assembly history of the VGS\_31 system in detail, we have obtained a multiwavelength data set. To investigate the 
low surface brightness features, we performed deep B and R band imaging, and to derive star formation properties and star formation history we have used deep $\rm{H_{\alpha}}$ 
and UV imaging. Gas morphology, kinematics and molecular hydrogen content have been investigated using 21\-cm HI and CO observations.
  
This paper is organized as follows: Section 2 describes the observations, data analysis and the derivation of star formation properties for VGS\_31.
Section 3 gives the main observational results such as star formation rate (SFR) properties, gas content and the morphology. In Section 4 we discuss
the results. Finally, in Section 5 we speculate on the nature and the importance of VGS\_31.

\begin{table}
 \caption{Observing summary for VGS\_31}
\begin{tabular}{ c c  }
Telescope & Observation details     \\
 \hline
\hline
INT   & B \& R band imaging \\
MDM 2.4m   & $\rm{H_{\alpha}}$ imaging \\
WHT   & Long slit spectrum  \\
WSRT  & 21cm (HI) imaging  \\
IRAM  & CO(1-0) observations \\ 
GALEX & NUV \& FUV imaging  \\
\hline
 \label{table:1}
\end{tabular}
\end{table}

\section{Observations and Data Reduction}

VGS\_31 has been observed in several wavelengths with various telescopes between 2009 and 2012. B \& R band imaging has been gathered with the Isaac Newton 
Telescope (INT) at La Palma using the Wide Field Camera (WFC). Long slit spectra were obtained at the William Herschel Telescope (WHT) using the Intermediate dispersion Spectrograph and 
Imaging System (ISIS). Narrow band $\rm{H_{\alpha}}$ imaging
has been done using the Hiltner Telescope at the Michigan-Dartmouth-MIT Observatory (MDM). Near UV (NUV) and far UV (FUV) images have been taken from the Nearby Galaxy Atlas (NGA)
of GALEX. Radio observations in the 21\-cm HI line were performed with the Westerbork Synthesis Radio Telescope (WSRT) and CO(1-0) emission spectra have been obtained
with IRAM. A summary of the observational studies is given in Table~\ref{table:1}. Parameters, quantitative results such as SFRs, HI and molecular hydrogen and stellar masses are presented in 
Table~\ref{table:2}, Table~\ref{table:3}, Table~\ref{table:4} and Table~\ref{table:5}, respectively.

\begin{deluxetable}{cccccccccccc}
\centering
\tabletypesize{\footnotesize}
\setlength{\tabcolsep}{0.02in} 
\tablewidth{0pt}
\tablecaption{Parameters for VGS\_31
\label{table:2}}
\tablenum{2}
\tablehead{\colhead{Name} & \colhead{SDSS ID} & \colhead{ra} & \colhead{dec} & \colhead{ z } & \colhead{$m_{r}$} & \colhead{$M_{r}$} & \colhead{$g - r$} & \colhead{$m_{B}$} & \colhead{$M_{B}$} & \colhead{$\delta$} & \colhead{} \\ 
\colhead{} & \colhead{} & \colhead{(J2000)} & \colhead{(J2000)} & \colhead{} & \colhead{} & \colhead{} & \colhead{} & \colhead{} & \colhead{} & \colhead{} & \colhead{} } 

\startdata
VGS\_31a & J131606.19+413004.2  & 13:16:06.19 & +41:30:04.25 & 0.021  &14.75 & -20.01 & 0.32 & 14.633 & -20.194 & -0.64  \\
VGS\_31b & J131614.69+412940.0 & 13:16:14.69 & +41:29:40.05 & 0.021  &14.38 & -20.38 & 0.50 & 14.632 & -20.193 & -0.64 \\
VGS\_31c & J131559.18+412955.9& 13:15:59.18 & +41:29:55.96 & 0.021  &16.78 & -17.98 & 0.21 & 16.735 & -18.093 & -0.64 \\
\enddata

\tablecomments{Column 1: Galaxy names. Column 2: SDSS IDs of the galaxies. Column 3 \& 4: Right ascensions and declinations. Column 5: Spectrophotometric redshifts. Column 6 \& 7: Apparent and 
absolute r magnitudes drawn from the SDSS DR7 and corrected for galactic extinction. Column 8: $g - r$ colors drawn from the apparent model magnitudes as measured by the SDSS DR7. 
Column 9 \& 10: Apparent  and absolute B magnitudes derived from INT B band imaging and corrected for galactic extinction. Column 11: $\delta$ gives the filtered density contrast
at $\rm{R_{\textit{f}}} = 1 h^{-1} Mpc$ as described in \cite{kreckel2011}.}
\end{deluxetable}

\subsection{HI imaging}

We have imaged the HI in VGS\_31 as part of the VGS project, the details of which are described in \cite{kreckel2012}. 
Observations were done with the Westerbork Synthesis Radio Telescope (WSRT) in the maxi-short configuration providing an angular 
resolution of $19^{\prime\prime} \times 32^{\prime\prime}$. The 36$^\prime$ full width half maximum of the WSRT primary beam is
sufficient to cover the entire VGS\_31 system in a single pointing. We observed 512 channels within a total bandwidth of 10 MHz, 
giving a Hanning smoothed velocity resolution of 8.6 km s$^{-1}$. Images for this paper were made with natural weighting to maximize 
sensitivity and CLEANed down to 0.5 mJy beam$^{-1}$ ($\sim$1 $\sigma$), reaching column density sensitivities of 2 $\times$ 10$^{19}$
 cm$^{-2}$.

\subsection{Broad band imaging and photometric calibration}

We used the Wide Field Camera (WFC) at the 2.1m INT for imaging in both B and R bands with Harris B and R filters. Total exposure times were
2400 second for B and 1800 seconds for the R band, spread over 4 exposures for the purpose of dithering and facilitating cosmic ray detection. 
Standard star fields were observed each night for the photometric calibration. Flat field exposures were taken at twilight at the beginning and/or end 
of each night. The data have been reduced using the standard IRAF\footnote{http://iraf.noao.edu/} and Photom Data Reduction Package 
(STARLINK)\footnote{http://star-www.rl.ac.uk/} procedures for CCD imaging. All the optical images were trimmed and overscanned followed by bias subtraction 
and flat fielding. After that all images from each filter were aligned and median combined. The same procedures have been followed for the standard star 
observations.

\subsection{$\rm{H_{\alpha}}$ imaging and photometric calibration}

$\rm{H_{\alpha}}$ imaging has been done with the Echelle CCD in direct mode at the 2.4 m Hiltner Telescope. A redshifted $\rm{H_{\alpha}}$ filter 
centered at 6693 $\mbox{\AA}$  has been used. To provide a measure of the continuum, R band imaging has been performed for each object. The total integration times for 
$\rm{H_{\alpha}}$ and for the continuum have been spread over 3 exposures for the purpose of dithering and for facilitating cosmic ray detection. 
Spectrophotomeric calibration stars have been chosen either from \cite{massey1988} or \cite{oke1990}. 

After performing the standard CCD reduction steps described in section 2.1, each combined $\rm{H_{\alpha}}$ image has been divided by 600 and R band
continuum image by 120 in order to normalize them to 1 second. The mean has been calculated for an empty region in each image and the ratio of these means
has been taken as the scaling factor for scaling the continuum image before subtraction from the $\rm{H_{\alpha}}$ image. Photometric calibration of the final $\rm{H_{\alpha}}$ images has
been performed following the steps described in \cite{gavazzi2006} and the references therein. Corrections for the atmospheric extinction and the airmass have been performed
in the standard way, where each spectrophotometric calibration star observation has been fitted using airmass and instrumental magnitudes to get the atmospheric extinction
coefficient. 

\textit{Contribution of the [N II] line} to the observed flux has been estimated using the expression from \cite{kennicutt2008} and has been subtracted from the $\rm{H_{\alpha}}$
flux.

\textit{Correction for the foreground extinction} has been derived from  Balmer decrements, following the recipes in \cite{calzetti2000} and \cite{dominguez2012}.the $\rm{H_{\alpha}}/\rm{ H_{\beta}}$ ratio
ratios have been obtained from the MPA-JHU catalog for the SDSS DR7\footnote{The MPA-JHU catalog is publicaly available and may be downloaded at http://www.mpa-garching.mpg.de/SDSS/DR7/archive}
measured through 3$''$ fibers. We have calculated E(B-V)
from these $\rm{H_{\alpha}}/\rm{ H_{\beta}}$ ratios, using the reddening curve from \cite{calzetti2000} to
obtain the corresponding extinction. By using Balmer decrements we correct the foreground extinction along the entire line of sight including
galactic extinction. Here an important point is that this correction assumes that the $\rm{H_{\alpha}}/\rm{ H_{\beta}}$ ratio is the same throughout the 
entire $\rm{H_{\alpha}}$ emission region. We checked this assumption using our WHT long slit spectra which measure the $\rm{H_{\alpha}}/\rm{ H_{\beta}}$ 
ratios along slits which covers the emission along both the minor and the major optical axis. In Table~\ref{table:3}, we give both the corrected and uncorrected 
$\rm{H_{\alpha}}$ fluxes.

\subsubsection{SFR from $\rm{H_{\alpha}}$ imaging}

Star formation rates (SFR) from the $\rm{H_{\alpha}}$ emission have been calculated following the conversion from \cite{kennicutt2009} where we used the SFR
conversion factor based on a ``Kroupa'' initial mass function \citep{kroupa2003}:
\begin{equation}
SFR (H_{\alpha}) \left[ M_{\odot} yr^{-1} \right] = 5.4 \times 10^{-42} \times L (H_{\alpha}), \nonumber
\end{equation}
where $ L (H_{\alpha}) $ is the luminosity, calculated as  
\begin{equation}
L(H_{\alpha}) \left [ erg\;s^{-1} \right ]= 4 \; \pi \; D^{2} (3.086 \times 10^{24})^{2} \; I(H_{\alpha}), \nonumber
\end{equation}
where D is the distance to the galaxy in Mpc and $I(H_{\alpha})$ is foreground extinction corrected flux. 

A detailed description of the $\rm{H_{\alpha}}$ photometry, foreground extinction correction and star formation derivation will be given in a
separate paper.

  \begin{figure*}
\begin{minipage}{180mm}
 \begin{center}
  \centering \includegraphics[scale = 0.3]{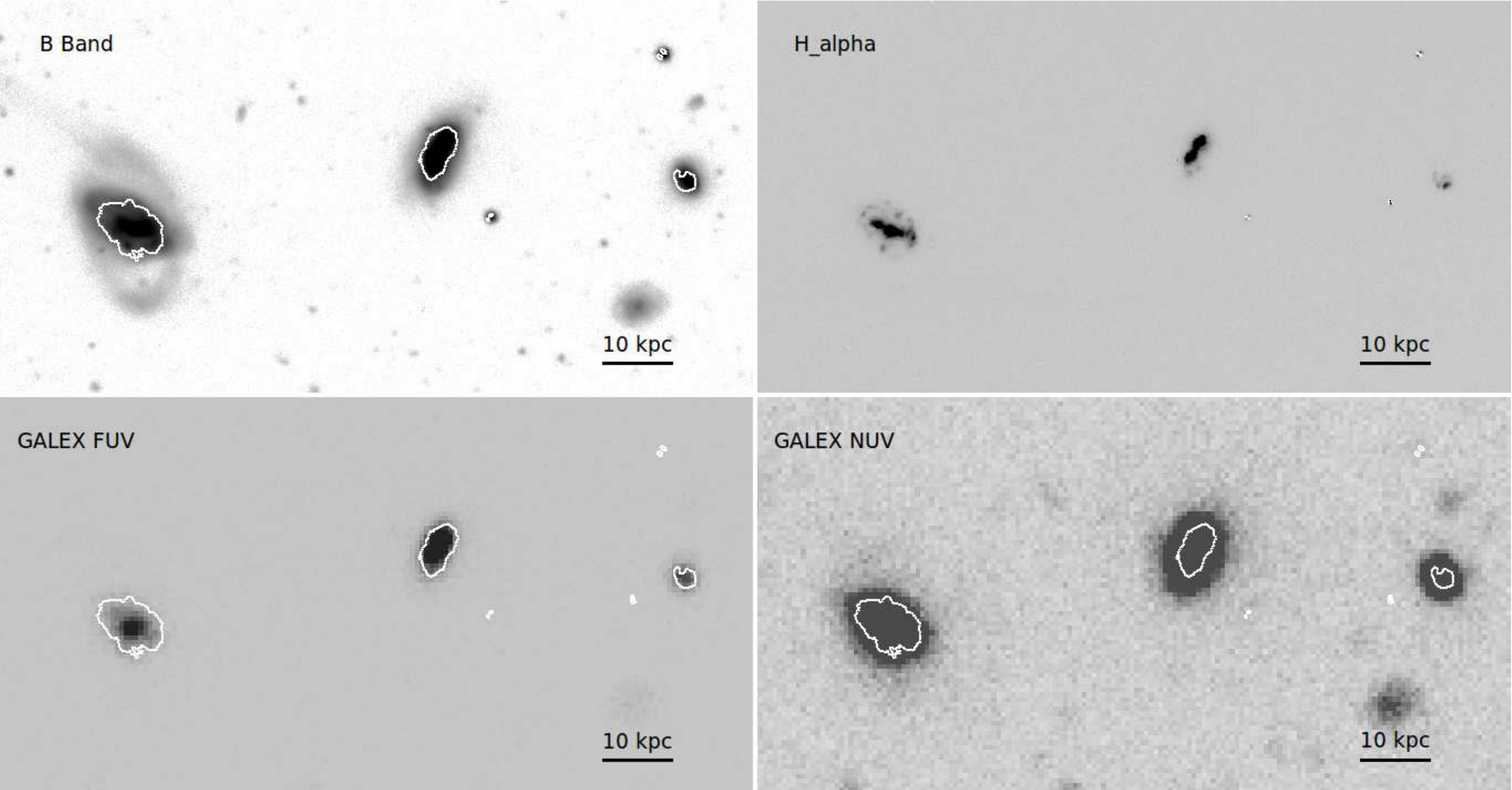}
 \end{center}
  \caption{VGS\_31 in four wavelength regimes. From top-left to bottom-right: a) INT B band image. b) MDM $\rm{H_{\alpha}}$ continuum-subtracted image
 c) GALEX Far UV image, d) GALEX Near UV image. All the images are shown at the same physical scale. The white contours on each image indicate the 
corresponding $\rm{ H_{\alpha}} $ emission regions. Note that there is no $\rm{ H_{\alpha}} $ emission in the ring or in the tail of VGS\_31b. Also 
note that the $\rm{ H_{\alpha}}$ emission is confined to the central parts of the galaxies, mostly in the bar structures of VGS\_31a and VGS\_31b.}
\label{figure:5}
\end{minipage}
 \end{figure*} 

\begin{figure*}
  \centering \includegraphics[scale = 0.5,angle=270]{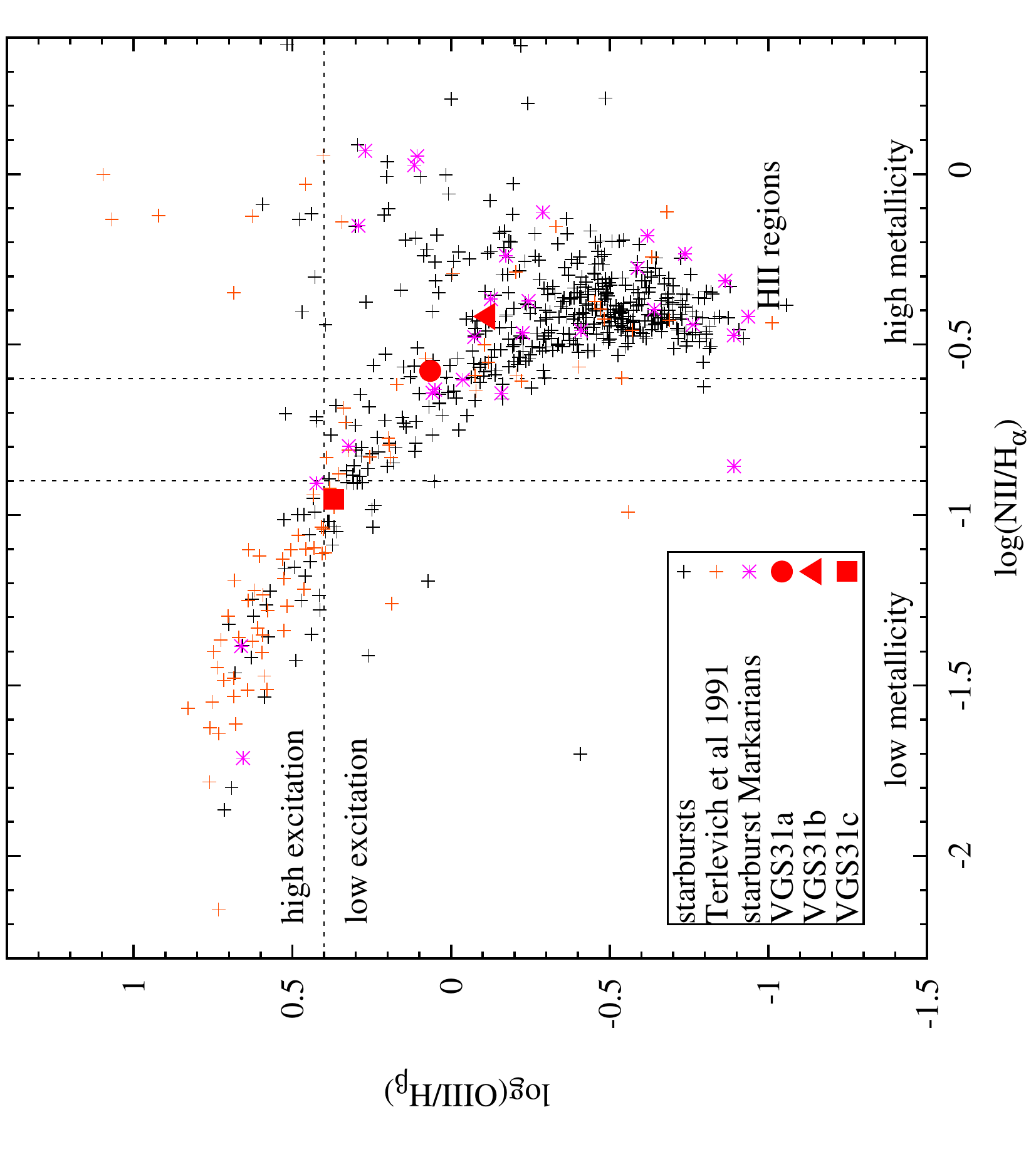}
  \caption{BPT diagram. The BPT diagram shows the ratio of emission line fluxes of [O III]/$\rm{H_{\beta}}$ to [N II]/$\rm{H_{\alpha}}$. The comparison sample of 
emission-line galaxies has been constracted from the emission-line galaxy sample of \cite{terlevich1991}, galaxies defined as starbursts in SIMBAD and Markarian 
galaxies defined as starbursts in \cite{coziol2003}. This diagram is adapted from \cite{raimann2000}.}
\label{figure:6}
\end{figure*} 

\subsection{Spectroscopy}

We used the Intermediate dispersion Spectrograph and Imaging System (ISIS) at the 4.2 m WHT at La Palma to take high resolution spectra. The R1200R grating in the
red arm and the R600B grating in the blue arm have been used, giving resolutions of respectively 0.026 nm/pixel over 620-720 nm and 0.045 nm/pixel
over 360-540 nm. CuNeAr lamp exposures were taken for wavelength calibrations. All the reduction has been performed using IRAF. In addition to the basic 
steps explained above, the illumination function along the slit has been determined via sky and lamp flats. Flux calibration has been done using 
spectrophotometric standard stars. 

Figure~\ref{figure:7} shows the $\rm{H_{\alpha}}$ emission along the optical major and minor axis of VGS\_31a and VGS\_31b. In this figure, we compare 
the $\rm{H_{\alpha}}$ emission line profiles with the $\rm{H_{\alpha}}$ images of the corresponding regions on the galaxies.

\subsection{GALEX UV observations}

GALEX NUV and FUV data have been obtained from the Nearby Galaxy Atlas (NGA), taken in 2004 with an exposure time of 3754 seconds in the NUV and 3002 seconds in the FUV.
VGS\_31 falls as a background object in the corresponding science frame. The data was calibrated on the basis of the GALEX pipelines.

\subsubsection{SFR from NUV}

The SFR has been calculated from the GALEX NUV and FUV luminosities and corrected for internal dust attenuation
following the method outlined in \cite{schiminovich2010}.
\begin{equation}
 SFR =\frac{L_{UV}f_{UV}(young)10^{0.4A_{UV}}}{\eta_{UV}},\nonumber
\end{equation}
where $L_{UV}$ is the luminosity in $erg\; s^{-1} Hz^{-1}$, $f_{UV}(young)$ is the fraction of light that originates
in young stellar populations, $\eta_{UV}$ is the conversion factor between UV luminosity and recent-past-
averaged star formation rate and $A_{UV}$ is the dust attenuation.

To determine the values of $A_{FUV}$ and $A_{NUV}$ we assume $f_{UV}(young)$ = 1 and $\eta_{UV}$ = $10^{28.165}$  \citep{schiminovich2010}. 

\subsection{CO(1-0) emission line observations}

CO(1-0) line observations have been carried out with the IRAM 30-m telescope at Pico Veleta, Spain in 2011. We used the Eight Mixer Receiver (EMIR) to
observe simultaneously the CO(1-0) (rest frequency, 115.271 GHz) and the CO(2-1) emission line (rest frequency, 222.8118 GHz) with a resolution of 5km/s. The full width half
maximum (FWHM) is, respectively, $\sim$22$''$ and $\sim$11$''$ at the two frequencies. The Wideband Line Multiple Autocorrelator (WILMA) and the Fast Fourier Transform Spectrometer 
(FTS) were used as backends. WILMA and FTS cover a channel width of 4 GHz with 2MHz and 195 KHz resolution, respectively. The observations were 
carried out in wobbler switching mode with a frequency of 1Hz and a throw of 120 $''$. The data were reduced with the CLASS software. 

\begin{figure}
  \centering \includegraphics[scale = 0.17]{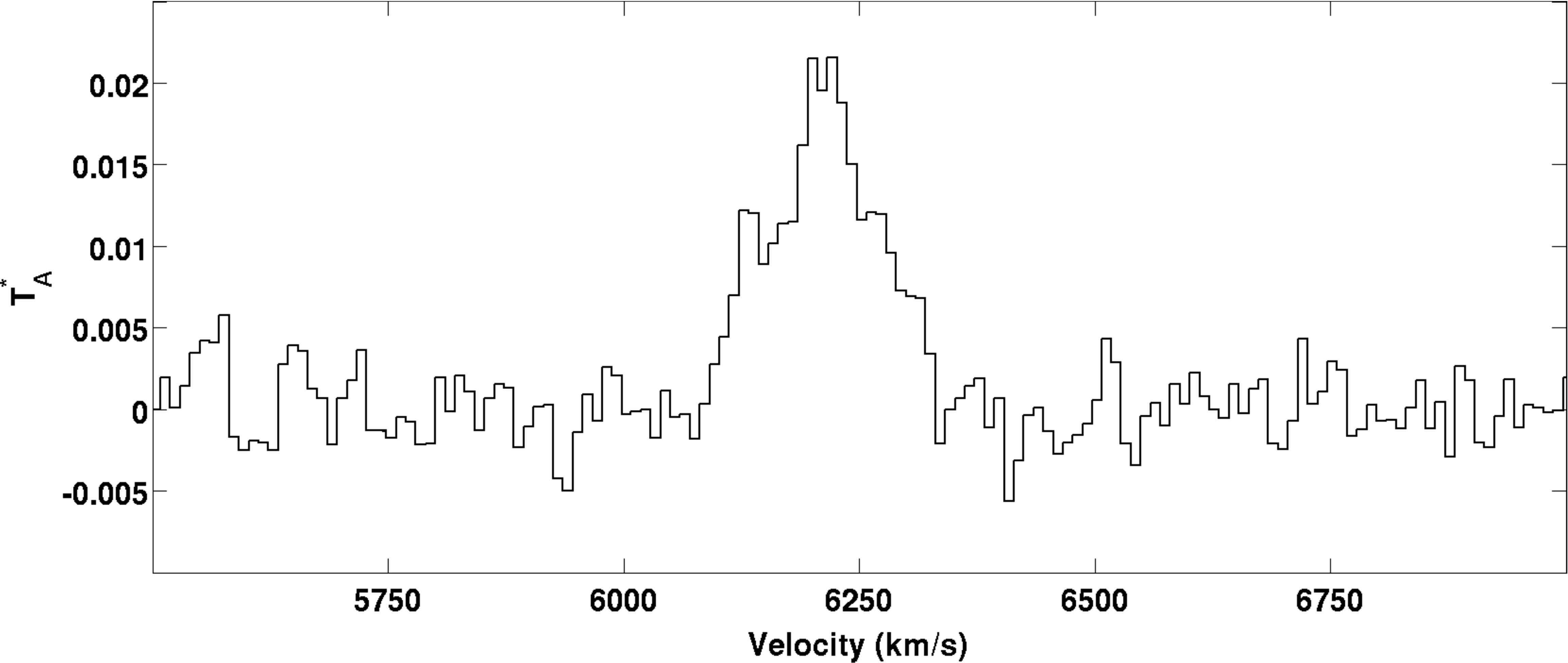}
  \caption{IRAM spectrum of VGS\_31b in the 115.271 GHz. The intensity scale is in $T^{*}_{A}$, in Kelvin.}
\label{figure:7}
\end{figure}

We have detected CO(1-0) emission from VGS\_31b and the profile is shown 
in Figure~\ref{figure:7}. VGS\_31b was not detected in CO(2-1). The detection limit for CO(2-1) is $\sim$ 1 $K \; km\;s^{-1}$. 
This is consistent with a normal CO(1-0)/CO(2-1) ratio, provided the CO(2-1) emission is distributed over the CO(1-0) beam area.

The data for VGS\_31a was not usable because of an error in the focus setting. VGS\_31c has not been 
observed.

\subsubsection{CO Luminosity and Mass}

IRAM spectra are expressed in terms of antenna temperature ($T^{*}_{A}$). In order to convert it to a flux density we adopted the expressions 
following \cite{costagliola2011} and \cite{saintonge2011}:

\begin{equation}
S_{CO}[Jy] = T^{*}_{A} [K] \times 3.906 \times \frac{F_{eff}}{\eta_{A}},\nonumber
\end{equation}
where $F_{eff}$ is the forward efficiency and $\eta_{A}$ is the antenna efficiency. The total flux is then derived by integrating the observed CO 
profile over velocity.

From the total flux we have calculated the luminosity as follows:

\begin{equation}
L'_{CO} = 3.25 \times 10^{7} \; S_{CO} \; \nu_{obs}^{-2} \; D_{L}^{2} \; (1+z)^{-3}, \nonumber
\end{equation} 
where $L'_{CO}$ is the luminosity, $\nu_{obs}$ is the observed frequency in units of GHz, $D_{L}$ is the luminosity distance in Mpc and z is redshift. Finally molecular 
hydrogen  mass $M_{H_{2}}$ is calculated as:

\begin{equation} 
M_{H_{2}} = L'_{CO} \times \alpha_{CO},\nonumber
\end{equation} 
where $\alpha_{CO}$ is the Galactic conversion factor of $3.2 M_{\odot} \; [(K\; km \;sec^{-1}pc^{2})^{-1}]$ which does not include a correction for the 
presence of Helium. 

\begin{figure*}
 \centering \includegraphics[scale = 0.15]{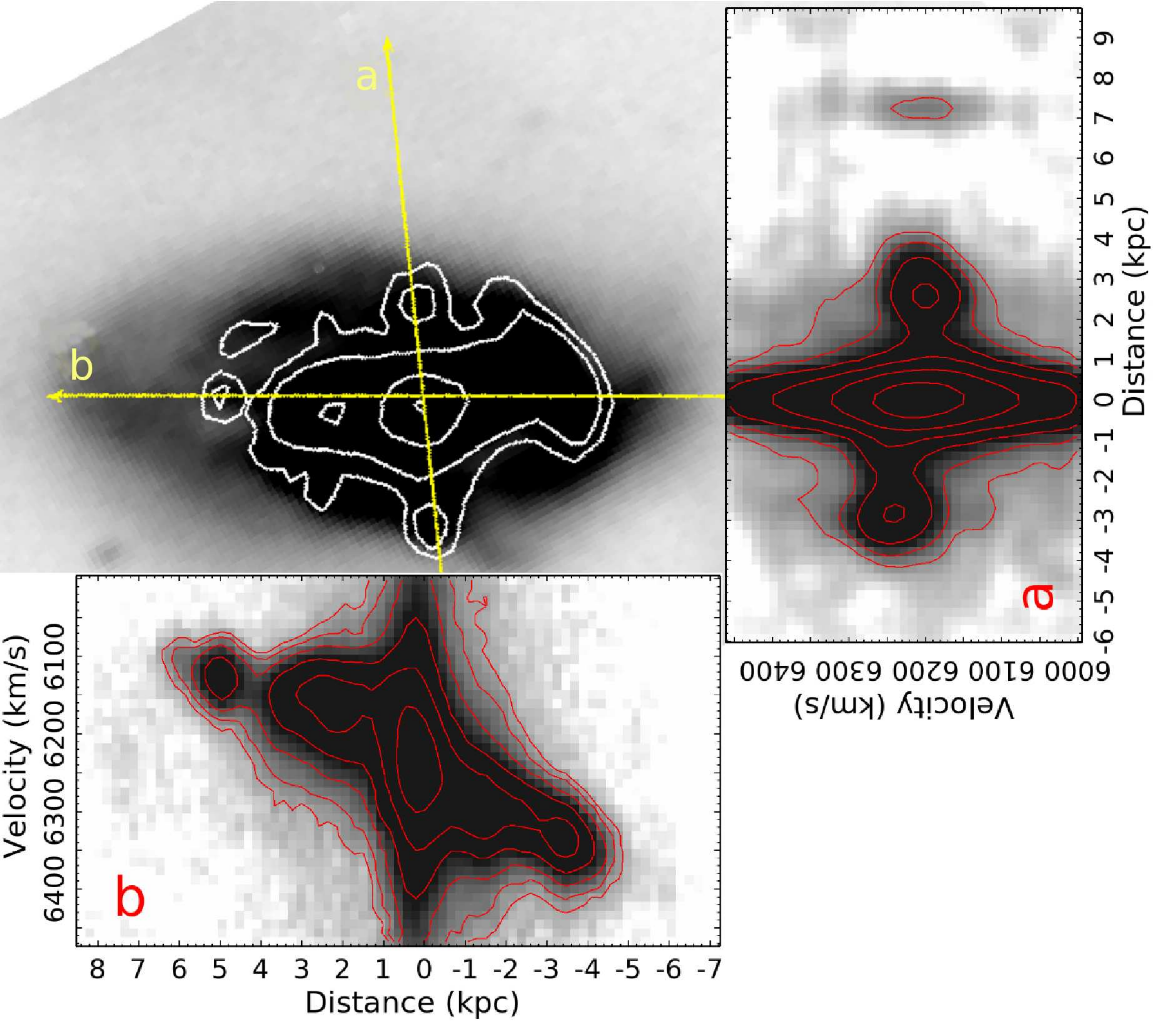}
 \centering \includegraphics[scale = 0.15]{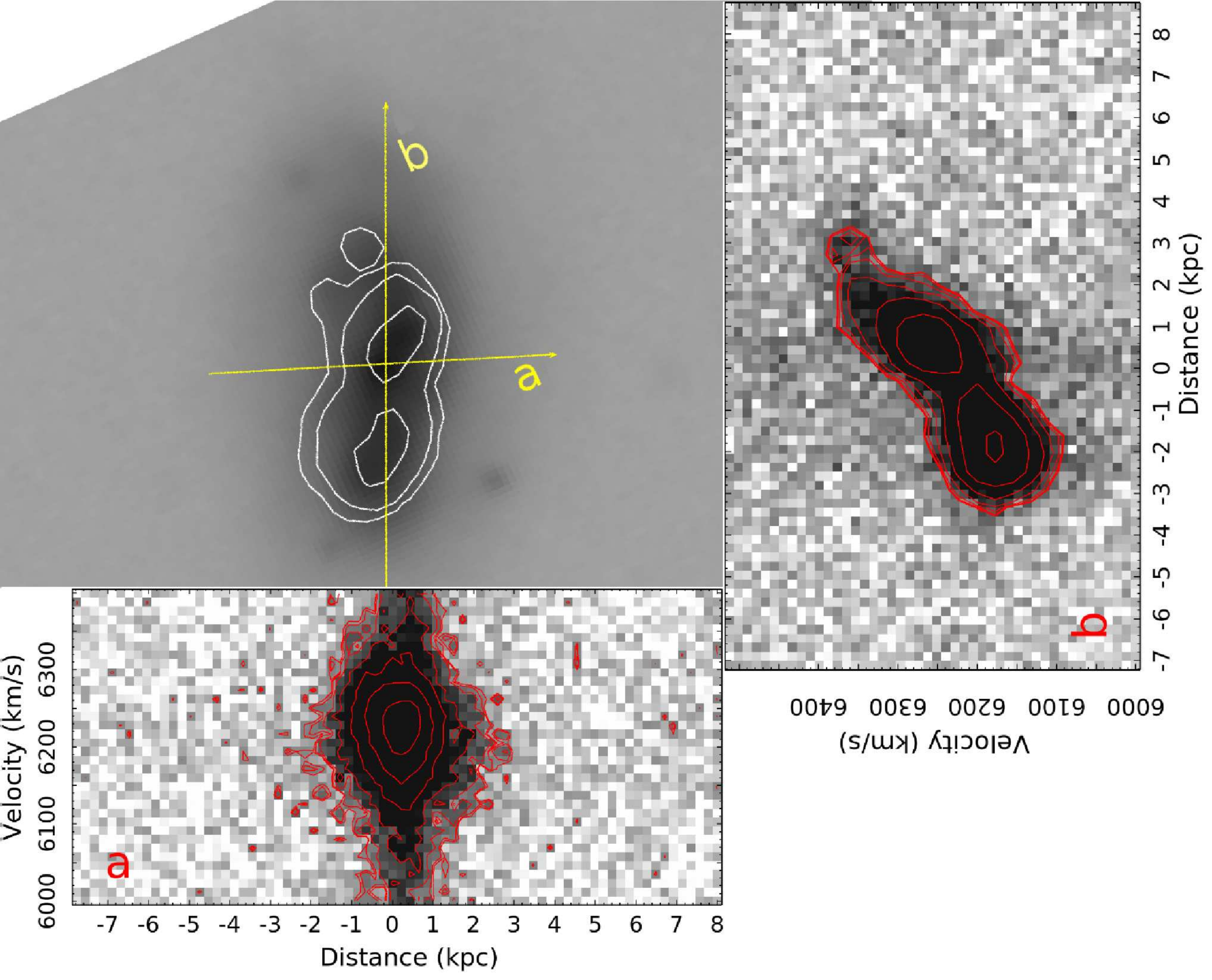}
  \caption{$\rm{H_{\alpha}}$ major and minor axis spectra and images for VGS\_31b (left) and VGS\_31a (right). In the two figures
 above,  the $\rm{H_{\alpha}}$ emission line spectra along the optical major and minor axis of VGS\_31b and VGS\_31a are shown on the same scale as their 
 optical images. In both figures, the optical images of the galaxies are overlaid by white $\rm{H_{\alpha}}$ imaging contours. On top of these, the 
yellow lines indicate
the positions of the slits. \textit{b} represents the slit position along the optical major axis and \textit{a} represents the slit position along
the optical minor axis. The optical images of the galaxies and the spectra are scaled in such a way that the physical sizes of the line profiles 
matches with the real size of the corresponding $\rm{H_{\alpha}}$ emission region within $\sim$ 0.6 kpc. Each diagram showing optical spectra have axes 
of relative distance from the centre (in kpc) vs. heliocentric radial velocity (in km/s). }
\label{figure:8}
\end{figure*} 

\begin{deluxetable}{ c c c c c }
\centering
\tablewidth{0pt}
\tablecaption{$\rm{H_{\alpha}}$ fluxes and luminosities
\label{table:3}}
\tablenum{3}
\tablehead{\colhead{Galaxy} & \colhead{log[$F(H_{\alpha}+ NII)$]} & \colhead{log[$[F(H_{\alpha})$]} & \colhead{log[$I(H_{\alpha})$]} & \colhead{$L(H_{\alpha})$} \\ 
\colhead{} & \colhead{} & \colhead{} & \colhead{} & \colhead{$(erg \;s^{-1})$}  } 

\startdata
VGS\_31a  & -12.622 & -12.755 & -12.304$\mp$0.076 &  4.7 $\times 10^{41}$  \\
VGS\_31b  & -12.345 & -12.478 & -11.913$\mp$0.104 &  1.16$\times 10^{42}$ \\
VGS\_31c  & -13.604 & -13.667 & -13.527$\mp$0.019 &  2.8 $\times 10^{40}$  \\
\enddata

\tablecomments{Column 1: Galaxy name. Column 2: Measured $\rm{H_{\alpha}}$ flux including NII lines. Column 3: $\rm{H_{\alpha}}$ flux corrected for NII deblending.
Column 4: Extinction corrected $\rm{H_{\alpha}}$ flux. Column 5: $\rm{H_{\alpha}}$ luminosity derived from the extinction corrected $\rm{H_{\alpha}}$
flux.}
\end{deluxetable}

\begin{deluxetable}{ c c c c c c }
\centering
\tablewidth{0pt}
\tablecaption{Star formation properties.
\label{table:4}}
\tablenum{4}
\tablehead{\colhead{Galaxy}& \colhead{$SFR_{\alpha}$} & \colhead{$SFR_{UV}$} & \colhead{$S\_SFR_{\alpha}$} & \colhead{$SFR_{\alpha}/M_{HI}$} &  \colhead{$SFR_{\alpha}/M_{H_{2}}$} \\
\colhead{} & \colhead{$(M_{\odot} yr^{-1})$} & \colhead{$(M_{\odot} yr^{-1})$} & \colhead{$(10^{-8} M_{\odot} yr^{-1})$} & \colhead{$(10^{-8} M_{\odot} yr^{-1})$} & \colhead{$(10^{-8} M_{\odot} yr^{-1})$} }

\startdata
VGS\_31a & 2.5  & 1.4  & 0.07  & 0.13 & - \\
VGS\_31b & 6.3  & 2.2  & 0.06  & 0.43 & 0.7  \\
VGS\_31c & 0.15 & 0.22 & 0.05  & 0.09 & -   \\
\enddata

\tablecomments{Column 1: Galaxy names. Column 2: Star formation measured from $\rm{H_{\alpha}}$ flux luminosities. Column 3: Star formation calculated from NUV 
Column 4: Specific star formation rate. Column 5: Star formation rate per HI mass. All three galaxies have similar $S\_SFR_{\alpha}$'s while their $SFR_{\alpha}/M_{HI}$ are different.}                                                    
\end{deluxetable}

 \begin{deluxetable}{c c c c c c c}
\centering
\tablewidth{0pt}
\tablecaption{Stellar, HI and molecular masses and CO(1-0) luminosities for VGS31.
\label{table:5}}
\tablenum{5}
\tablehead{\colhead{Galaxy} & \colhead{$M_{*}$} & \colhead{$M_{HI}$} & \colhead{$L_{CO}$} & \colhead{$M_{H_{2}}$} & \colhead{$M_{H_{2}}/M_{*}$} & \colhead{$M_{H_{2}}/M_{HI}$}    \\
\colhead{} & \colhead{$(10^{8} M_{\odot})$} & \colhead{$(10^{8} M_{\odot})$} & \colhead{$(10^{8}$ K km $s^{-1} pc^2)$} & \colhead{$(10^{8} M_{\odot})$} & \colhead{} & \colhead{} } 

\startdata
VGS\_31a & 35.1   & 19.89 $\pm$ 2.90 & -    &   - & -     & - \\
VGS\_31b & 105.31 & 14.63 $\pm$ 1.97 & 2.8 & 8.9 & 0.085 & 0.608 \\
VGS\_31c &  2.92  & 1.66 $\pm$ 0.95  & -    & -   &  -    & - \\
\enddata

\tablecomments{Stellar masses are taken from the publicly available MPA-JHU catalog for SDSS DR7 which were derived using fits to the broad-band \textit{ugriz} photometry.
VGS31 galaxies are all spectroscopic targets in SDSS DR7 with 3$''$ fiber spectra in their central regions and these spectra have been analyzed and included in this catalog. }                                                    
\end{deluxetable}
 
\section{Results}

Combining the data from the diverse set of observations presented in Section 2, the following picture on the nature of VGS\_31 emerges.

VGS\_31 consists of three galaxies located in a void, VGS\_31a, b and c (Figure~\ref{figure:1}). A careful study of the HI properties suggests that
the galaxies are embedded in an elongated HI cloud (Figure~\ref{figure:3} and 4). In addition, both the optical and the HI observations suggest strong 
interactions in the system (Figure~\ref{figure:3}). This emerging picture
tells us that we may be dealing with two different processes. One is the assembly of a filamentary structure in a void, the other is an interaction
between the galaxies. We will describe these two main results separately.

VGS\_31 exhibits a very peculiar structure in HI. The HI column density map and the position-velocity (PV) diagram (Figure~\ref{figure:3} and 4) present two views elucidating the nature of the system;
\textit{i) the filamentary structure} and \textit{ii) its kinematic properties}. 

\subsection{Filamentary structure}

The PV diagram seen on the right hand side of Figure~\ref{figure:4}, is created by taking one single slice through the HI data cube, from the 
far east to the far 
west of the VGS\_31. In the HI column density map (Figure~\ref{figure:3}) galaxies appear to be lined up in a HI filament. The whole HI envelope
extends over $\sim$ 120 kpc on the sky. The PV diagram shows that the 
three galaxies are at about the same systemic velocity. Also the extreme ends of the filament are at the same velocities suggesting a single, coherent filamentary structure. 

\subsection{Kinematic properties}

The kinematics along this perceived filamentary structure is, however, rather complex. We will examine this, galaxy by galaxy, going from east to west in Figure~\ref{figure:4}. 

\textit{VGS\_31b}: The optical tail has a HI counterpart, showing almost no velocity gradient. There is a slight offset between this HI tail and its 
optical counterpart. It is kinematically connected to the inner disk and the ring of VGS\_31b.
The spatial resolution is insufficient to distinguish the gas associated with the ring from the gas associated with the disk. Therefore, we can not say whether the HI tail is connected to the
inner disk or the ring. The galaxy disk and the
ring exhibit rotation with a
velocity spread of 325 $\rm{km \;s^{-1}}$. The $\rm{H_{\alpha}}$ spectrum (Figure~\ref{figure:8}) shows the rotation 
along the optical major axis. In the PV diagram, the velocity gradient is not as steep because the slice has not 
been exactly placed along the optical major axis but rather towards the tail. West of the disk, we see a HI bridge between VGS\_31b and VGS\_31a. 
Close to VGS\_31b the HI in this bridge has a velocity dispersion of $\sim$ 50 $km\;s^{-1}$.
In \textit{VGS\_31b}, there is almost no velocity gradient in the $\rm{H_{\alpha}}$ emission along the 
optical minor axis whereas we see a significant gradient along the optical major axis. 

\textit{VGS\_31a}: However, halfway to VGS\_31a the velocity dispersion becomes significantly larger,$\sim$ 125 $km\;s^{-1}$. This behaviour is repeated at the other side of VGS\_31a as well.
The HI bridge has not been detected in the optical or UV. The gas associated with VGS\_31a shows broad velocity range. However, we don't have the spatial resolution to map the HI kinematics
in detail. As we move away from the galaxy towards the
west, we continue to see HI with a broad velocity dispersion. The $\rm{H_{\alpha}}$ spectrum (Figure~\ref{figure:8}), has much higher resolution and shows a large kinematic asymmetry. 
 
\textit{VGS\_31c}: Very little HI is detected between VGS\_31a and VGS\_31c. A clear HI connection is however visible in Figure~\ref{figure:4}.
The velocity dispersion of the HI bridge becomes again narrower as we approach VGS\_31c.
As for the bridge between VGS\_31b and VGS\_31a, there is no stellar counterpart for this HI bridge either. 
The HI resolution is not enough to determine the detailed gas kinematics of VGS\_31c.

\subsection{Structure of the galaxies}

In addition to HI, optical data complement the picture of VGS\_31's complex dynamics. We will continue to describe the peculiar features of the 
system going from east to west.

\textit{VGS\_31b:} As seen from Figure~\ref{figure:1}, VGS\_31b is, optically, the most disturbed galaxy. It has a one sided tail, curved towards 
the north east. There is a ring like
structure around the disk. The tail and the disk seem to be connected. The first zoomed image shows the inner disk. 
The second zoomed image displays the bar positioned asymmetrically in the disk and the bright central part. The $\rm{H_{\alpha}}$ emission 
of VGS\_31b shows clear rotation along the
optical major axis. In addition there is a very fast rotating inner structure, about 1 kpc in extent with a velocity width of $\sim$ 500 km/s. Along 
the minor axis a minor velocity diference 
is seen, indicative of non-circular motions along the bar (Figure~\ref{figure:8}). Also, all star formation is concentrated in this central region 
where the bar is present. 
VGS\_31b has been detected in CO(1-0). The velocity of the CO(1-0)
peak is at $\sim$ 6200 $km\;s^{-1}$, which is similar to the velocity of the HI.

\textit{VGS\_31a:} The zoom-in image (Figures~\ref{figure:1}) of VGS\_31a shows that it is slightly disturbed and the bar like structure in the 
center overlaps with the $\rm{H_{\alpha}}$ emission region (Figures~\ref{figure:5}). Like VGS\_31b, rotation is along the optical major axis. 
However, the $\rm{H_{\alpha}}$ emission line profile along the major axis is quite irregular. 

\textit{VGS\_31c:} This is the smallest of the three galaxies without any significant morphological irregularities (Table~\ref{table:5}). As in the other two galaxies, the star formation is
confined to the central part of the disk.

\subsection{Star formation properties}

The $\rm{H_{\alpha}}$ and UV results show that all three galaxies exhibit recent star formation activity concentrated in their central parts 
(Figure~\ref{figure:5} and Table~\ref{table:4} ). The tail and the ring of VGS\_31b are not detected in $\rm{H_{\alpha}}$ or UV. The same is true for the HI bridges between VGS\_31b, VGS\_31a and VGS\_31c. Among the three galaxies, VGS\_31b
has significantly higher $\rm{SFR_{\alpha}}$ (Table~\ref{table:4}). We also emphasize that the small $D_{n}(4000)$ break values are indications for
young stellar populations in the central part of the galaxies.

From the SDSS spectra we may determine the location of the VGS\_31 members in a Baldwin, Phillips \& Terlevich (BPT) diagram \citep{baldwin1981} shown in Figure~\ref{figure:6}. For this diagram, emission lines
of all galaxies, including VGS\_31, have been extracted from the SDSS DR7 spectral database of 3" fiber apertures. In this diagram, both VGS\_31a and 
VGS\_31b are located in the HII zone together with the other starburst galaxy samples. Their SDSS spectra
have strong Balmer emission lines with large equivalent widths and blue continua. VGS\_31a and VGS\_31b are located inside the HII/ starburst region
in the BPT diagram while VGS\_31c is placed inside the normal star forming zone.

Stellar masses of the three galaxies range from $\rm{3 \times 10^{8} M_{\odot}}$ to $\rm{1.06 \times 10^{10} M_{\odot}}$ (Table~\ref{table:5}). In contrast to the difference in their stellar masses,
they have similar $\rm{S\_SFR_{\alpha}}$. It is worth nothing that their $\rm{S\_SFR_{\alpha}}$ and $\rm{SFR_{\alpha}/M_{HI}}$ are significantly above the median of 
those of the ALFALFA average density sample \citep{kreckel2012}, indicating enhanced star formation in comparison with galaxies of similar mass and 
gas content.

\section{Discussion}

VGS\_31 is a  peculiar system through which we may be witnessing the ongoing growth of three galaxies along a filament inside a void. Here we discuss 
possible scenarios for the evolution of the
VGS\_31 galaxies in the void including gas accretion from an intra-void filament, interactions and merging. These processes are difficult to disentangle
from one another on the basis of the observational material presented. Yet we will discuss the observed phenomena in the context of all of
these processes and indicate which we consider most important in each of the individual galaxies.

First we will discuss different interaction scenarios for VGS\_31a and VGS\_31b using the observational results presented above. VGS\_31b is the most 
eye-catching member in the system.
A one sided tidal tail and a ring like structure are clearly visible in the optical (Figure~\ref{figure:1}). This morphology suggests a minor merging 
incident with a low mass galaxy. A satellite galaxy
could have left the tail and formed the ring by wrapping the disk of VGS\_31b \citep{mihos1994,mihos1996,duc2011}. It is less likely that the tail has 
been caused by tidal
interaction between VGS\_31b and VGS\_31a. As the mass ratio of these systems is 3 to 1 one would expect greater damage to the disks and more prominent 
tails and counter-tail features as 
usually seen in  tidal interactions and major mergers \citep{hibbard1996}. The disks of VGS\_31b and VGS\_31a are not destroyed as expected in minor 
mergers \citep{schweizer2000,delgado2010,duc2011}. 

VGS\_31b has a bar, visible in close up images in Figure~\ref{figure:1}. The bar is pronounced in $\rm{H_{\alpha}}$ as well (Figure~\ref{figure:5}). 
The kinematics of the
$\rm{H_{\alpha}}$ (Figure~\ref{figure:8}) indicate a fast rotating inner structure and evidence for streaming motions characteristic for a bar. 
In addition, VGS\_31b is a starburst Markarian
galaxy having enhanced star formation in its central part overlapping with the bar (Figure~\ref{figure:5} and~\ref{figure:8}, Table~\ref{table:4}). 
The detection of CO and location of 
VGS\_31b in the BPT diagram support the starburst picture, presumably the consequence of gas accretion into its central part as seen in many galaxy 
mergers and
interactions \citep{mihos1994,mihos1996,duc2011}. There is an offset between the stellar and the gas component of the tail (Figure~\ref{figure:3}). 
This is also observed in some tidally
interacting systems \citep{mihos2000} and may be due to different initial distributions of both components or additional processes that act on one 
component and not on the other 
\citep[see][for a recent review]{duc2011}. On the other hand, in most of the tidal tails in mergers, ongoing star formation is observed 
\citep{hibbard1996,neff2004}, in the case of
VGS\_31b, however, no ongoing star formation has been detected neither in the tail nor in the ring (Figure~\ref{figure:5}).

VGS\_31a, however, shows different characteristics. It has no visible tails, but the internal kinematics is disturbed (see the $\rm{H_{\alpha}}$ 
kinematics in Figure~\ref{figure:7}) and the HI
shows that gas east and west of the
galaxy exhibits a large spread in velocity as if there is a corotating halo filled with HI rather than tidal features with simple kinematics 
(Figure~\ref{figure:4}). Like VGS\_31b, VGS\_31a
is a starburst galaxy with enhanced star formation in its central part (Figure~\ref{figure:5} and~\ref{figure:6}, Table~\ref{table:4}). There are two 
possibilities for the mechanism which could cause the enhanced star formation and the morphological disturbance: \textit{i)} VGS\_31a could be in 
interaction with VGS\_31b and this may cause the disturbence in its morphology and results in the irregular $\rm{H_{\alpha}}$ emission line profile. 
\textit{ii)} Instead of accreting gas from VGS\_31b, VGS\_31a may experience steady gas infall from the intergalactic medium, presumably from the 
structure outlined by the location of the galaxies and the HI `connecting' them (Figure~\ref{figure:4}). 
This could cause the enhanced star formation as in the first scenario and better explains the broad velocity range of the HI surrounding VGS\_31a. 
One could argue that if VGS\_31a exhibits such accretion from an LSS filament, then one would also expect VGS\_31b to show the same characteristics. 
On the other hand it is conceivable that a minor merger, as suggested by the tail
and ring could have disrupted the accretion process. Detailed simulations of such scenarios are required to test the validity of
these scenarios.

VGS\_31c, the smalles of the three galaxies, has enhanced star formation
and its HI shows the charecteristics of interactions, albeit at very
low signal to noise. It is difficult in this case to determine conclusively
the process(es) responsible for the HI structure and enhanced star formation.

The most exciting result is that we may be witnessing the assembly of structure within a void, and the birth process of 
galaxies in such a desolate area. It does fit into the theoretically 
expected buildup of voids and galaxies therein. Voids evolve in a hierarchical fashion, leaving planar and filamentary substructure 
within the emerging voids
 \citep{dubinski1993,weykamp1993,sahni1994,sheth2004,aragon2013} and (Rieder et al. 2013 (subm.)). 
The question is whether we see here the manifestation of this process. VGS\_31 could be a density enhancement within an
 underlying dark matter filament. The complication
is that in addition to accretion of material from a LSS filament the
galaxies also suffer from tidal interactions and minor merging. As for the cold flow accretion scenarios we can only make a very rough estimate because 
we don't 
know exactly the orientation of 
the system. If we take the size of the HI filament and the velocity gradient from east to west then we can estimate a rough accretion timescale of at 
least a gigayear.
An important next step will be to use advanced simulations with gas and
star formation to see whether the scenario proposed here does indeed
take place in voids.

Finally, it is of interest to note that VGS\_31 is not the only case for a filamentary structure and cold flow accretion in our Void Galaxy Survey.
We have discovered at least two other cases which seem to exhibit this hierarchical structure 
formation. In addition to another 
filamentary galaxy configuration, VGS\_38 \citep{kreckel2011}, we also found a polar disk galaxy, VGS\_12 \citep{stanonik2009}. VGS\_38 is a 
system of chain galaxies which 
share the same HI. VGS\_12 is located right at the center
of a tenuous wall between two large voids. It has a polar HI disk much more extended than its stellar disk. The polar disk has no stellar 
counterpart or any ongoing star formation. This galaxy is a candidate for the cold flow accretion. 

Individually these galaxies and VGS\_31 are unusual systems, however taken collectively they show that the void environment is an extremely
interesting site for understanding galaxy formation and evolution.

\section*{Acknowledgments}

We wish to thank the anonymous referee for interesting and helpful comments. RvdW and BB are grateful to Steven Rieder and Marius Cautun 
for many helpful discussions and insightful thoughts. BB wishes to thank Reynier Peletier for his help in spectral 
analysis and the staff of the IRAM observatory for their tremendous help in conducting our observations and IGN (Spain). 
This work was supported in part by the National Science Foundation under grant no. 1009476 to Columbia University. We are grateful 
for support from Da Vinci Professorship at the Kapteyn Institute. 
The Isaac Newton Telescope and The William Herschel Telescope are operated on the island of La Palma by the Isaac Newton Group 
in the Spanish Observatorio del Roque de los Muchachos of the Instituto de Astrofísica de Canarias. MDM observatory is located on the
southwest ridge of Kitt Peak, home of the Kitt Peak National Observatory, Tucson, Arizona. The Observatory is owned and operated 
by a consortium of five universities: the University of Michigan, Dartmouth College, the Ohio State University, Columbia University, 
and Ohio University. We acknowledge KPNO for the use of their $\rm{H_{\alpha}}$ filters. The Westerbork Synthesis Radio Telescope 
is operated by the ASTRON (Netherlands Institute for Radio Astronomy) with support from the Netherlands Foundation for Scientific 
Research (NWO). IRAM is supported by INSU/CNRS (France), MPG (Germany) and has benefited from research funding from the European 
Community's Seventh Framework Programme.


\begin{thebibliography}{68}
\expandafter\ifx\csname natexlab\endcsname\relax\def\natexlab#1{#1}\fi

\bibitem[{{Arag{\'o}n-Calvo} {et~al.}(2010{\natexlab{a}}){Arag{\'o}n-Calvo},
  {Platen}, {van de Weygaert}, \& {Szalay}}]{aragon2010a}
{Arag{\'o}n-Calvo}, M.~A., {Platen}, E., {van de Weygaert}, R., \& {Szalay},
  A.~S. 2010{\natexlab{a}}, \apj, 723, 364

\bibitem[{{Arag{\'o}n-Calvo} \& {Szalay}(2013)}]{aragon2013}
{Arag{\'o}n-Calvo}, M.~A., \& {Szalay}, A.~S. 2013, \mnras, 428, 3409

\bibitem[{{Arag{\'o}n-Calvo} {et~al.}(2010{\natexlab{b}}){Arag{\'o}n-Calvo},
  {van de Weygaert}, \& {Jones}}]{aragon2010b}
{Arag{\'o}n-Calvo}, M.~A., {van de Weygaert}, R., \& {Jones}, B.~J.~T.
  2010{\natexlab{b}}, \mnras, 408, 2163

\bibitem[{{Baldwin} {et~al.}(1981){Baldwin}, {Phillips}, \&
  {Terlevich}}]{baldwin1981}
{Baldwin}, J.~A., {Phillips}, M.~M., \& {Terlevich}, R. 1981, \pasp, 93, 5

\bibitem[{{Bond} {et~al.}(1996){Bond}, {Kofman}, \& {Pogosyan}}]{bond1996}
{Bond}, J.~R., {Kofman}, L., \& {Pogosyan}, D. 1996, \nat, 380, 603

\bibitem[{{Calzetti} {et~al.}(2000){Calzetti}, {Armus}, {Bohlin}, {Kinney},
  {Koornneef}, \& {Storchi-Bergmann}}]{calzetti2000}
{Calzetti}, D., {Armus}, L., {Bohlin}, R.~C., {Kinney}, A.~L., {Koornneef}, J.,
  \& {Storchi-Bergmann}, T. 2000, \apj, 533, 682

\bibitem[{{Ceccarelli} {et~al.}(2006){Ceccarelli}, {Padilla}, {Valotto}, \&
  {Lambas}}]{ceccar2006}
{Ceccarelli}, L., {Padilla}, N.~D., {Valotto}, C., \& {Lambas}, D.~G. 2006,
  MNRAS, 373, 1440

\bibitem[{{Colberg} {et~al.}(2005{\natexlab{a}}){Colberg}, {Krughoff}, \&
  {Connolly}}]{colberg2005a}
{Colberg}, J.~M., {Krughoff}, K.~S., \& {Connolly}, A.~J. 2005{\natexlab{a}},
  \mnras, 359, 272

\bibitem[{{Colberg} {et~al.}(2005{\natexlab{b}}){Colberg}, {Sheth}, {Diaferio},
  {Gao}, \& {Yoshida}}]{colberg2005b}
{Colberg}, J.~M., {Sheth}, R.~K., {Diaferio}, A., {Gao}, L., \& {Yoshida}, N.
  2005{\natexlab{b}}, \mnras, 360, 216

\bibitem[{{Colless} {et~al.}(2003){Colless}, {Peterson}, {Jackson}, {Peacock},
  {Cole}, {Norberg}, {Baldry}, {Baugh}, {Bland-Hawthorn}, {Bridges}, {Cannon},
  {Collins}, {Couch}, {Cross}, {Dalton}, {De Propris}, {Driver}, {Efstathiou},
  {Ellis}, {Frenk}, {Glazebrook}, {Lahav}, {Lewis}, {Lumsden}, {Maddox},
  {Madgwick}, {Sutherland}, \& {Taylor}}]{colless2003}
{Colless}, M., {et~al.} 2003, ArXiv Astrophysics e-prints

\bibitem[{{Costagliola} {et~al.}(2011){Costagliola}, {Aalto}, {Rodriguez},
  {Muller}, {Spoon}, {Mart{\'{\i}}n}, {Per{\'e}z-Torres}, {Alberdi},
  {Lindberg}, {Batejat}, {J{\"u}tte}, {van der Werf}, \&
  {Lahuis}}]{costagliola2011}
{Costagliola}, F., {et~al.} 2011, \aap, 528, A30

\bibitem[{{Coziol}(2003)}]{coziol2003}
{Coziol}, R. 2003, \mnras, 344, 181

\bibitem[{{Davis} {et~al.}(1983){Davis}, {Huchra}, \& {Latham}}]{huchra1983}
{Davis}, M., {Huchra}, J., \& {Latham}, D. 1983, in IAU Symposium, Vol. 104,
  Early Evolution of the Universe and its Present Structure, ed. G.~O. {Abell}
  \& G.~{Chincarini}, 167--172

\bibitem[{{Dom{\'{\i}}nguez} {et~al.}(2012){Dom{\'{\i}}nguez}, {Siana},
  {Henry}, {Scarlata}, {Bedregal}, {Malkan}, {Atek}, {Ross}, {Colbert},
  {Teplitz}, {Rafelski}, {McCarthy}, {Bunker}, {Hathi}, {Dressler}, {Martin},
  \& {Masters}}]{dominguez2012}
{Dom{\'{\i}}nguez}, A., {et~al.} 2012, ArXiv e-prints

\bibitem[{Dubinski {et~al.}(1993)Dubinski, da~Costa, Goldwirth, Lecar, \&
  Piran}]{dubinski1993}
Dubinski, J., da~Costa, L.~N., Goldwirth, D.~S., Lecar, M., \& Piran, T. 1993,
  ApJ, 410, 458

\bibitem[{{Duc} \& {Renaud}(2011)}]{duc2011}
{Duc}, P.-A., \& {Renaud}, F. 2011, ArXiv e-prints

\bibitem[{{Einasto} {et~al.}(2011){Einasto}, {Suhhonenko}, {H{\"u}tsi}, {Saar},
  {Einasto}, {Liivam{\"a}gi}, {M{\"u}ller}, {Starobinsky}, {Tago}, \&
  {Tempel}}]{einasto2011}
{Einasto}, J., {et~al.} 2011, \aap, 534, A128

\bibitem[{Furlanetto \& Piran(2006)}]{furlanetto2006}
Furlanetto, S., \& Piran, T. 2006, MNRAS, 366, 467

\bibitem[{{Gavazzi} {et~al.}(2006){Gavazzi}, {Boselli}, {Cortese}, {Arosio},
  {Gallazzi}, {Pedotti}, \& {Carrasco}}]{gavazzi2006}
{Gavazzi}, G., {Boselli}, A., {Cortese}, L., {Arosio}, I., {Gallazzi}, A.,
  {Pedotti}, P., \& {Carrasco}, L. 2006, \aap, 446, 839

\bibitem[{{Gottl{\"o}ber} {et~al.}(2003){Gottl{\"o}ber}, {{\L}okas}, {Klypin},
  \& {Hoffman}}]{gottloeb2003}
{Gottl{\"o}ber}, S., {{\L}okas}, E.~L., {Klypin}, A., \& {Hoffman}, Y. 2003,
  MNRAS, 344, 715

\bibitem[{{Grogin} \& {Geller}(1999)}]{grogin1999}
{Grogin}, N.~A., \& {Geller}, M.~J. 1999, \aj, 118, 2561

\bibitem[{{Grogin} \& {Geller}(2000)}]{grogin2000}
---. 2000, \aj, 119, 32

\bibitem[{{Hibbard} \& {van Gorkom}(1996)}]{hibbard1996}
{Hibbard}, J.~E., \& {van Gorkom}, J.~H. 1996, \aj, 111, 655

\bibitem[{{Hoyle} \& {Vogeley}(2002)}]{hoyle2002a}
{Hoyle}, F., \& {Vogeley}, M.~S. 2002, \apj, 566, 641

\bibitem[{{Hoyle} {et~al.}(2002){Hoyle}, {Vogeley}, \& {Gott}}]{hoyle2002b}
{Hoyle}, F., {Vogeley}, M.~S., \& {Gott}, III, J.~R. 2002, \apj, 570, 44

\bibitem[{{Huchra} {et~al.}(2012){Huchra}, {Macri}, {Masters}, {Jarrett},
  {Berlind}, {Calkins}, {Crook}, {Cutri}, {Erdo{\v g}du}, {Falco}, {George},
  {Hutcheson}, {Lahav}, {Mader}, {Mink}, {Martimbeau}, {Schneider},
  {Skrutskie}, {Tokarz}, \& {Westover}}]{huchra2012}
{Huchra}, J.~P., {et~al.} 2012, \apjs, 199, 26

\bibitem[{{Karachentseva} {et~al.}(1999){Karachentseva}, {Karachentsev}, \&
  {Richter}}]{karachentseva1999}
{Karachentseva}, V.~E., {Karachentsev}, I.~D., \& {Richter}, G.~M. 1999, A\&AS,
  135, 221

\bibitem[{{Kennicutt} {et~al.}(2008){Kennicutt}, {Lee}, {Funes}, {Sakai}, \&
  {Akiyama}}]{kennicutt2008}
{Kennicutt}, Jr., R.~C., {Lee}, J.~C., {Funes}, Jos{\'e}~G., S.~J., {Sakai},
  S., \& {Akiyama}, S. 2008, \apjs, 178, 247

\bibitem[{{Kennicutt} {et~al.}(2009){Kennicutt}, {Hao}, {Calzetti},
  {Moustakas}, {Dale}, {Bendo}, {Engelbracht}, {Johnson}, \&
  {Lee}}]{kennicutt2009}
{Kennicutt}, Jr., R.~C., {et~al.} 2009, \apj, 703, 1672

\bibitem[{{Kreckel} {et~al.}(2012){Kreckel}, {Platen}, {Arag{\'o}n-Calvo}, {van
  Gorkom}, {van de Weygaert}, {van der Hulst}, \& {Beygu}}]{kreckel2012}
{Kreckel}, K., {Platen}, E., {Arag{\'o}n-Calvo}, M.~A., {van Gorkom}, J.~H.,
  {van de Weygaert}, R., {van der Hulst}, J.~M., \& {Beygu}, B. 2012, \aj, 144,
  16

\bibitem[{{Kreckel} {et~al.}(2011){Kreckel}, {Platen}, {Arag{\'o}n-Calvo}, {van
  Gorkom}, {van de Weygaert}, {van der Hulst}, {Kova{\v c}}, {Yip}, \&
  {Peebles}}]{kreckel2011}
{Kreckel}, K., {et~al.} 2011, \aj, 141, 4

\bibitem[{{Kroupa} \& {Weidner}(2003)}]{kroupa2003}
{Kroupa}, P., \& {Weidner}, C. 2003, \apj, 598, 1076

\bibitem[{{Kuhn} {et~al.}(1997){Kuhn}, {Hopp}, \& {Elsaesser}}]{kuhn1997}
{Kuhn}, B., {Hopp}, U., \& {Elsaesser}, H. 1997, \aap, 318, 405

\bibitem[{{Mart{\'{\i}}nez-Delgado} {et~al.}(2010){Mart{\'{\i}}nez-Delgado},
  {Gabany}, {Crawford}, {Zibetti}, {Majewski}, {Rix}, {Fliri},
  {Carballo-Bello}, {Bardalez-Gagliuffi}, {Pe{\~n}arrubia}, {Chonis}, {Madore},
  {Trujillo}, {Schirmer}, \& {McDavid}}]{delgado2010}
{Mart{\'{\i}}nez-Delgado}, D., {et~al.} 2010, \aj, 140, 962

\bibitem[{{Massey} {et~al.}(1988){Massey}, {Strobel}, {Barnes}, \&
  {Anderson}}]{massey1988}
{Massey}, P., {Strobel}, K., {Barnes}, J.~V., \& {Anderson}, E. 1988, \apj,
  328, 315

\bibitem[{{Mihos}(2000)}]{mihos2000}
{Mihos}, C. 2000, ArXiv Astrophysics e-prints

\bibitem[{{Mihos} \& {Hernquist}(1994)}]{mihos1994}
{Mihos}, J.~C., \& {Hernquist}, L. 1994, \apjl, 425, L13

\bibitem[{{Mihos} \& {Hernquist}(1996)}]{mihos1996}
---. 1996, \apj, 464, 641

\bibitem[{{Neff} {et~al.}(2004){Neff}, {Thilker}, {Hibbard}, {Seibert}, {Gil de
  Paz}, {Schiminovich}, {Martin}, {Rich}, {Madore}, \& {Bianchi}}]{neff2004}
{Neff}, S.~G., {et~al.} 2004, in Bulletin of the American Astronomical Society,
  Vol.~36, American Astronomical Society Meeting Abstracts, 1385

\bibitem[{{Oke}(1990)}]{oke1990}
{Oke}, J.~B. 1990, \aj, 99, 1621

\bibitem[{{Patiri} {et~al.}(2006{\natexlab{a}}){Patiri}, {Betancort-Rijo},
  {Prada}, {Klypin}, \& {Gottl{\"o}ber}}]{patiri2006a}
{Patiri}, S.~G., {Betancort-Rijo}, J.~E., {Prada}, F., {Klypin}, A., \&
  {Gottl{\"o}ber}, S. 2006{\natexlab{a}}, MNRAS, 369, 335

\bibitem[{{Patiri} {et~al.}(2006{\natexlab{b}}){Patiri}, {Prada}, {Holtzman},
  {Klypin}, \& {Betancort-Rijo}}]{patiri2006b}
{Patiri}, S.~G., {Prada}, F., {Holtzman}, J., {Klypin}, A., \&
  {Betancort-Rijo}, J. 2006{\natexlab{b}}, MNRAS, 372, 1710

\bibitem[{{Peebles}(2001)}]{peebles2001}
{Peebles}, P.~J.~E. 2001, \apj, 557, 495

\bibitem[{{Platen} {et~al.}(2007){Platen}, {van de Weygaert}, \&
  {Jones}}]{platen2007}
{Platen}, E., {van de Weygaert}, R., \& {Jones}, B.~J.~T. 2007, \mnras, 380,
  551

\bibitem[{{Popescu} {et~al.}(1997){Popescu}, {Hopp}, \&
  {Elsaesser}}]{popescu1997}
{Popescu}, C.~C., {Hopp}, U., \& {Elsaesser}, H. 1997, \aap, 325, 881

\bibitem[{{Portegies Zwart} {et~al.}(2010){Portegies Zwart}, {Ishiyama},
  {Groen}, {Nitadori}, {Makino}, {de Laat}, {McMillan}, {Hiraki}, {Harfst}, \&
  {Grosso}}]{portegies2010}
{Portegies Zwart}, S., {et~al.} 2010, IEEE Computer, v.43, No.8, p.63-70, 43,
  63

\bibitem[{{Raimann} {et~al.}(2000){Raimann}, {Storchi-Bergmann}, {Bica},
  {Melnick}, \& {Schmitt}}]{raimann2000}
{Raimann}, D., {Storchi-Bergmann}, T., {Bica}, E., {Melnick}, J., \& {Schmitt},
  H. 2000, \mnras, 316, 559

\bibitem[{{Rojas} {et~al.}(2004){Rojas}, {Vogeley}, {Hoyle}, \&
  {Brinkmann}}]{rojas2004}
{Rojas}, R.~R., {Vogeley}, M.~S., {Hoyle}, F., \& {Brinkmann}, J. 2004, \apj,
  617, 50

\bibitem[{{Rojas} {et~al.}(2005){Rojas}, {Vogeley}, {Hoyle}, \&
  {Brinkmann}}]{rojas2005}
---. 2005, \apj, 624, 571

\bibitem[{{Sahni} {et~al.}(1994){Sahni}, {Sathyaprakah}, \&
  {Shandarin}}]{sahni1994}
{Sahni}, V., {Sathyaprakah}, B.~S., \& {Shandarin}, S.~F. 1994, ApJ, 431, 20

\bibitem[{{Saintonge} {et~al.}(2011){Saintonge}, {Kauffmann}, {Kramer},
  {Tacconi}, {Buchbender}, {Catinella}, {Fabello}, {Graci{\'a}-Carpio}, {Wang},
  {Cortese}, {Fu}, {Genzel}, {Giovanelli}, {Guo}, {Haynes}, {Heckman},
  {Krumholz}, {Lemonias}, {Li}, {Moran}, {Rodriguez-Fernandez}, {Schiminovich},
  {Schuster}, \& {Sievers}}]{saintonge2011}
{Saintonge}, A., {et~al.} 2011, \mnras, 415, 32

\bibitem[{{Schaap} \& {van de Weygaert}(2000)}]{schaap2000}
{Schaap}, W.~E., \& {van de Weygaert}, R. 2000, \aap, 363, L29

\bibitem[{{Schiminovich} {et~al.}(2010){Schiminovich}, {Catinella},
  {Kauffmann}, {Fabello}, {Wang}, {Hummels}, {Lemonias}, {Moran}, {Wu},
  {Giovanelli}, {Haynes}, {Heckman}, {Basu-Zych}, {Blanton}, {Brinchmann},
  {Budav{\'a}ri}, {Gon{\c c}alves}, {Johnson}, {Kennicutt}, {Madore}, {Martin},
  {Rich}, {Tacconi}, {Thilker}, {Wild}, \& {Wyder}}]{schiminovich2010}
{Schiminovich}, D., {et~al.} 2010, \mnras, 408, 919

\bibitem[{{Schweizer}(2000)}]{schweizer2000}
{Schweizer}, F. 2000, in Royal Society of London Philosophical Transactions
  Series A, Vol. 358, Astronomy, physics and chemistry of H$^{+}$$_{3}$, 2063

\bibitem[{{Shandarin} \& {Zeldovich}(1989)}]{shandarin1989}
{Shandarin}, S.~F., \& {Zeldovich}, Y.~B. 1989, Reviews of Modern Physics, 61,
  185

\bibitem[{{Sheth} \& {van de Weygaert}(2004)}]{sheth2004}
{Sheth}, R.~K., \& {van de Weygaert}, R. 2004, \mnras, 350, 517

\bibitem[{{Springel} {et~al.}(2006){Springel}, {Frenk}, \&
  {White}}]{springel2006}
{Springel}, V., {Frenk}, C.~S., \& {White}, S.~D.~M. 2006, \nat, 440, 1137

\bibitem[{{Stanonik} {et~al.}(2009){Stanonik}, {Platen}, {Arag{\'o}n-Calvo},
  {van Gorkom}, {van de Weygaert}, {van der Hulst}, \&
  {Peebles}}]{stanonik2009}
{Stanonik}, K., {Platen}, E., {Arag{\'o}n-Calvo}, M.~A., {van Gorkom}, J.~H.,
  {van de Weygaert}, R., {van der Hulst}, J.~M., \& {Peebles}, P.~J.~E. 2009,
  \apjl, 696, L6

\bibitem[{{Szomoru} {et~al.}(1996){Szomoru}, {van Gorkom}, {Gregg}, \&
  {Strauss}}]{szomoru1996}
{Szomoru}, A., {van Gorkom}, J.~H., {Gregg}, M.~D., \& {Strauss}, M.~A. 1996,
  \aj, 111, 2150

\bibitem[{{Terlevich} {et~al.}(1991){Terlevich}, {Melnick}, {Masegosa},
  {Moles}, \& {Copetti}}]{terlevich1991}
{Terlevich}, R., {Melnick}, J., {Masegosa}, J., {Moles}, M., \& {Copetti},
  M.~V.~F. 1991, \aaps, 91, 285

\bibitem[{{Tikhonov} \& {Karachentsev}(2006)}]{tikhonov2006}
{Tikhonov}, A.~V., \& {Karachentsev}, I.~D. 2006, ApJ, 653, 969

\bibitem[{{van de Weygaert} \& {Platen}(2011)}]{weyplaten2011}
{van de Weygaert}, R., \& {Platen}, E. 2011, International Journal of Modern
  Physics Conference Series, 1, 41

\bibitem[{{van de Weygaert} \& {Schaap}(2009)}]{weyschaap2009}
{van de Weygaert}, R., \& {Schaap}, W. 2009, in Lecture Notes in Physics,
  Berlin Springer Verlag, Vol. 665, Data Analysis in Cosmology, ed. V.~J.
  {Mart{\'{\i}}nez}, E.~{Saar}, E.~{Mart{\'{\i}}nez-Gonz{\'a}lez}, \& M.-J.
  {Pons-Border{\'{\i}}a}, 291--413

\bibitem[{van~de Weygaert \& van Kampen(1993)}]{weykamp1993}
van~de Weygaert, R., \& van Kampen, E. 1993, MNRAS, 263, 481

\bibitem[{{Wegner} \& {Grogin}(2008)}]{wegner2008}
{Wegner}, G., \& {Grogin}, N.~A. 2008, AJ, 136, 1

\bibitem[{{York} {et~al.}(2000){York}, {Adelman}, {Anderson}, {Anderson},
  {Annis}, {Bahcall}, {Bakken}, {Barkhouser}, {Bastian}, {Berman}, {Boroski},
  {Bracker}, {Briegel}, {Briggs}, {Brinkmann}, {Brunner}, {Burles}, {Carey},
  {Carr}, {Castander}, {Chen}, {Colestock}, {Connolly}, {Crocker}, {Csabai},
  {Czarapata}, {Davis}, {Doi}, {Dombeck}, {Eisenstein}, {Ellman}, {Elms},
  {Evans}, {Fan}, {Federwitz}, {Fiscelli}, {Friedman}, {Frieman}, {Fukugita},
  {Gillespie}, {Gunn}, {Gurbani}, {de Haas}, {Haldeman}, {Harris}, {Hayes},
  {Heckman}, {Hennessy}, {Hindsley}, {Holm}, {Holmgren}, {Huang}, {Hull},
  {Husby}, {Ichikawa}, {Ichikawa}, {Ivezi{\'c}}, {Kent}, {Kim}, {Kinney},
  {Klaene}, {Kleinman}, {Kleinman}, {Knapp}, {Korienek}, {Kron}, {Kunszt},
  {Lamb}, {Lee}, {Leger}, {Limmongkol}, {Lindenmeyer}, {Long}, {Loomis},
  {Loveday}, {Lucinio}, {Lupton}, {MacKinnon}, {Mannery}, {Mantsch}, {Margon},
  {McGehee}, {McKay}, {Meiksin}, {Merelli}, {Monet}, {Munn}, {Narayanan},
  {Nash}, {Neilsen}, {Neswold}, {Newberg}, {Nichol}, {Nicinski}, {Nonino},
  {Okada}, {Okamura}, {Ostriker}, {Owen}, {Pauls}, {Peoples}, {Peterson},
  {Petravick}, {Pier}, {Pope}, {Pordes}, {Prosapio}, {Rechenmacher}, {Quinn},
  {Richards}, {Richmond}, {Rivetta}, {Rockosi}, {Ruthmansdorfer}, {Sandford},
  {Schlegel}, {Schneider}, {Sekiguchi}, {Sergey}, {Shimasaku}, {Siegmund},
  {Smee}, {Smith}, {Snedden}, {Stone}, {Stoughton}, {Strauss}, {Stubbs},
  {SubbaRao}, {Szalay}, {Szapudi}, {Szokoly}, {Thakar}, {Tremonti}, {Tucker},
  {Uomoto}, {Vanden Berk}, {Vogeley}, {Waddell}, {Wang}, {Watanabe},
  {Weinberg}, {Yanny}, {Yasuda}, \& {SDSS Collaboration}}]{york2000}
{York}, D.~G., {et~al.} 2000, \aj, 120, 1579

\bibitem[{{Zel'dovich}(1970)}]{zeldovich1970}
{Zel'dovich}, Y.~B. 1970, \aap, 5, 84

\bibitem[{{Zitrin} \& {Brosch}(2008)}]{zitrin2008}
{Zitrin}, A., \& {Brosch}, N. 2008, \mnras, 390, 408

\end{thebibliography}

\end{document}